\documentclass[preprint,authoryear,11pt]{elsarticle}
\usepackage[english]{babel}

\journal{Studies in History and Philosophy of Modern Physics}

\newcommand{\cov}[1]{{#1}_\mu}

\newcommand{\con}[1]{{#1}^\mu}

\newcommand{\oneo}[1]{\frac{1}{#1}}

\newcommand{\pdt}[1]{\frac{\partial {#1}}{\partial t}}

\newcommand{\covgr}{\partial_\mu}
\newcommand{\congr}{\partial^\mu}

\newcommand{\vgrad}{\vec{\nabla}}

\newcommand{\dw}{\psi}

\newcommand{\mattwo}[4]{\left( \begin{array}{cc} {#1} & {#2} \\ {#3} & {#4} \end{array} \right)}

\begin{document}

\begin{frontmatter}

\title{ Surely You Must All be Joking: An Outsider's Critique of Quantum Physics }

\author{Randall C. O'Reilly}
\ead{randy.oreilly@colorado.edu}

\address{Department of Psychology and Neuroscience\\
  University of Colorado Boulder \\
  345 UCB\\
  Boulder, CO 80309}

\begin{abstract}
  A critique of the state of current quantum theory in physics is presented,
  based on a perspective outside the normal physics training.  From this
  perspective, the acceptance of quantum nonlocality seems unwarranted, and
  the fundamental assumptions that give rise to it in the first place seem
  questionable, based on the current status of the quantum theory of light.
  The relevant data can instead be accounted for using physically motivated
  local models, based on detailed properties of the experimental setups.  The
  semiclassical approach, particularly in the form of the fully coupled
  Maxwell-Dirac equations with a pure wave ontology, seems to provide a
  satisfying, local, paradox-free physical model of the quantum world, that
  appears consistent with known phenomena.  It is unclear why this approach is
  not pursued more vigorously in the field, given its clear potential to
  resolve all the conundrums that have perplexed generations of physicists.
\end{abstract}


\begin{keyword}
Quantum mechanics \sep Local realism \sep Bell's inequality \sep Maxwell-Dirac equations
\end{keyword}

\end{frontmatter}


\section{Introduction}

Sometimes an outsider can see things that those indoctrinated in a given field
have become blind to.  As a computational neuroscientist with a longstanding
interest in the fundamental laws of the universe, I have followed developments
in physics over the years.  I've read various lay accounts of the mysteries of
the quantum world and the quest for grand unified theories, including the
Feynman classics, the inscrutable book by Hawking, and the compelling books by
Brian Greene.  Recently, I dug extensively into the primary scientific
literature, in an attempt to gain clearer insight into the deeper mysteries
facing the field since the founding of quantum mechanics (QM) in the early
1900's.  I was sufficiently shocked by what I found, that I felt compelled to
write this outsider's critique.  It is written in a deliberately provocative
tone, in the hopes of stimulating people into further reflection about some
foundational issues in the way the field has developed.

The quantum world apparently exhibits a number of strange properties,
including randomness, complementarity, wave-particle duality, and nonlocality.
Virtually every major figure in the field has attested to the fundamental
incomprehensibility of this world, e.g., Feynman's famous claim that ``I think
I can safely say that nobody understands quantum mechanics.''  As I've delved
deeper into the primary literature, I can see why: the verbal descriptions of
quantum physics in introductory material are often completely at odds with the
actual mathematical and conceptual frameworks that experts actually use
\citep[e.g.,][]{Klassen11}, and these frameworks are obviously just {\em
  calculational tools}, rife with virtual, non-physical entities and
gratuitous non-localities.  But in this primary literature, I also found the
apparently neglected work of a number of physicists, that seems to paint an
entirely sensible and comprehensible alternative, physical model.

This physical model is based entirely on the ontology of waves, which is (to
my surprise) in fact the effective ontology of the vast majority of the
mathematics of QM \citep{Nikolic07}, despite the seemingly perverse continued
insistence on describing things in terms of particles.  Doing away with
particles entirely seems to resolve a large number of apparent paradoxes and
fundamental confusions.  To make this pure-waves viewpoint work, one still
needs to wrestle with a number of unsolved problems, but there are plausible
solutions to each of these problems, even with the tiny smattering of
attention they have received.  The prospects of obtaining a sane and
comprehensible quantum worldview would seem to be sufficient motivation to put
significant effort into solving these problems.

But there is one major roadblock for the purely wave-based approach, which
seems to have caused most physicists to write it off from the outset.  This
problem is the apparent nonlocality of the quantum word.  According to the
standard interpretation, two {\em entangled} particles can interact with each
other {\em instantaneously} at a distance, in principle even if they are at
opposite ends of the universe!  Despite many years and concerted effort, it
seems that nobody has been able to provide a convincing way out of this
conclusion, and it has reduced many a respectable physicist to producing long
rambling discussions that inevitably just seem to skirt the central issues,
and dress them up in new terminology (quantum information, consistent
histories, many worlds, relational, etc..).

However, I have yet to see anyone make the following argument, which questions
the primary assumptions upon which quantum nonlocality is based:
\begin{itemize}
\item Only quantum entanglement of light can demonstrate nonlocality, because
  entangled massive particles can always interact via speed-of-light
  mechanisms.
\item The QM description of light in terms of photons is a complete disaster
  from a conceptual point of view.  There hasn't been a proper first quantized
  description of a photon wave function until relatively recently, and the
  only viable proposal propagates according to the same Maxwell's equations as
  the classical electromagnetic (EM) field \citep{BialynickiBirula94,Sipe95}.
  The second quantized version of the photon in quantum electrodynamics (QED)
  is manifestly both nonlocal and nonphysical --- it is a pure harmonic
  standing wave in Fourier space, stretching in principle across the entire
  universe.  Indeed, QED is sufficiently underconstrained that it is not even
  clear within this framework if photons actually travel at the speed of light
  --- recent experiments confirm that in fact they do \citep{ZhangEtAl11}.
\item Thus, the idea that one can transparently derive a physically sensible
  prediction about the entangled behavior of photons within the QM framework
  seems rather suspect.  However, every treatment of the standard QM
  predictions for entanglement experiments that I've seen uses a simple
  abstract Hilbert space formalism with no attempt to actually derive some
  kind of detailed physical model of what is actually going on in the exact
  experiments being performed.  I genuinely have no precise idea what people
  are even referring to when they use the term ``two entangled photons'' ---
  certainly they cannot be referring literally to QED photons as standing
  waves stretching across the universe?
\item The {\em semiclassical} approach to understanding quantum phenomena,
  where light is treated classically (i.e., regular old Maxwell's equations),
  and all the strange quantum behavior is attributable to the atomic system,
  has been successful beyond anyone's wildest expectations
  \citep{JaynesCummings63,Jaynes73,Mandel76,Grandy91,GerryKnight05,MarshallSantos97}.
  Indeed, this semiclassical approach appears to be widely used in the field
  of quantum optics, preferred in many instances over QED because it provides
  a much more natural physical picture of what actually seems to be happening
  in the relevant experiments (QED can give highly accurate results, but can
  also easily produce nonsense if not used properly)
  \citep{GerryKnight05,RoychoudhuriRoy03}.  To date, there do not appear to be
  {\em any} quantum phenomena that {\em some} semiclassical model can't
  account for, including the recent photon statistics experiments
  (anticorrelation and antibunching)
  \citep{MarshallSantos88,MarshallSantos97}, which had been regarded as the
  strongest evidence in support of the photon concept, and against the
  semiclassical approach \citep{GraingerRogerAspect86,HongOuMandel87}.
\item {\em If it is true that Maxwell's equations provide a valid description
    of the quantum behavior of light, then there is no reason to believe that
    light should exhibit quantum entanglement!}  There is certainly nothing
  within Maxwell's equations that would support entanglement of two light
  waves that are moving away from each other at the speed of light.  Given the
  success of the semiclassical approach in accounting in principle for all
  known quantum optics phenomena, shouldn't that give people considerable
  pause in accepting nonlocal entanglement?  Have people really given sober
  consideration to the tradeoff between the various scepticisms about the
  semiclassical approach, weighed against the complete insanity of the
  orthodox view of quantum nonlocality?  Or has any will to question the
  insanity been completely drained in the years of conceptual turmoil and
  systematic brainwashing?
\item Meanwhile, there appears to be a relatively unquestioning acceptance of
  the empirical evidence for photon entanglement, in the face of what appear
  to be very strong limitations.  What are commonly marginalized as ``loopholes''
  seem instead like very plausible physical descriptions of the actual
  behavior of the polarization detectors used in these experiments.  There are
  several principled local models that reproduce the observed data quite
  accurately
  \citep{MarshallSantosSelleri83,MarshallSantos85,Thompson96,AdenierKhrennikov03,Santos05,AschwandenEtAl06,AdenierKhrennikov07},
  and one analysis that provides positive evidence that the main ``loophole''
  is in fact operative according to the recorded data
  \citep{AdenierKhrennikov07}.  Furthermore, some of the results have been
  discrepant with QM predictions, and yet no attempt to replicate in the right
  way has been attempted \citep{Santos04}.
\end{itemize}

One obvious reason that nobody questions the fundamental assumption that QM
applies to photons must be that the QM framework is so incredibly successful,
how could it possibly be wrong in this case?  But outside of the peculiar
Bell's inequality tests for quantum nonlocality, the vast majority of QM
physics (in the pure wave ontology at least), is entirely compatible with
locality (including all those supposedly quantum phenomena captured by the
semiclassical approach).  As we discuss later, it seems that the standard
mathematical framework for QM just doesn't allow one to express the behavior
of {\em formerly entangled} photons (absent a measurement or even
decoherence), but this may very well just be a limitation of this descriptive
framework, and the underlying physics could be different, without anyone ever
noticing except in this very strange case.

In summary, it seems entirely plausible that there is no strong reason to
believe that quantum physics is necessarily nonlocal.  The things you have to
``give up'' to make it local seem like a very small price to pay in the grand
scheme of things, and they feel like the true message that the quantum world
has been trying to tell us all this time (but people have been too obsessed
with particles to really see it): nature at its most fundamental level is made
entirely of waves.  Waves are intrinsically {\em contextual} --- what you
measure about them reflects properties of both the incoming wave and the
measurement device, and they are spatially distributed, not discretely
localized like particles.  These properties have also been mistakenly
interpreted as implying nonlocality.

Thus, in contrast to everything we have been told, it seems that there may in
fact be a perfectly sensible physical model of the quantum world.  This model
would be based on locally propagating wave dynamics, with different kinds of
waves (electromagnetic waves, electron waves, etc) interacting in potentially
complex ways.  This idea was originally espoused by Schr\"odinger, but
discarded for reasons that may no longer hold up \citep{Dorling87}.  One of the
major reasons for rejecting this idea was certainly the presumed nonlocality
of the quantum world.  Maybe it is time to take another look?

In the remainder of the paper, the above ideas are developed in greater depth,
starting with some general terminological and conceptual clarifications.

\section{Calculational Tools vs. Physical Models}

To establish some terminology and a conceptual framework for the discussion,
it is important to distinguish between two major categories of theory in
physics: {\em calculational tools} and {\em physical models}.  Calculational
tools are systematic frameworks that provide a convenient representation of
physical problems for computing predictions of experimental results, but the
central constructs of these tools need not (and typically do not) provide a
model of how physics is actually thought to operate in nature.  There are
often nonlocal abstractions, and many decisions that require expert human
judgment in configuring the computations.  Any physical framework that
contains ``virtual'' or other non-physical entities is {\em by definition} a
calculational tool.

In contrast, physical models realistically describe objective physical
processes, operating universally and {\em autonomously}, that give rise to the
observed physical phenomena.  The notion of autonomy provides a critical
distinction between the two kinds of frameworks: whereas calculational tools
typically require lots of expert knowledge of how to represent a given
physical situation, a physical model can just iteratively crank away without
any expert intervention, and accurately reproduce the known physics.  Perhaps
the epitome of a physical model is the cellular automaton, which captures
exactly this notion of a simple autonomous system cranking iteratively away,
and numerous people have argued is the most compelling overall framework for
fundamental physics
\citep{Ulam50,Gardner70,Zuse70,FredkinToffoli82,Fredkin90,Wolfram83,ToffoliMargolus90,Poundstone85,BialynickiBirula94,Meyer96}.
Calculational tools can typically produce results in one step, whereas
physical models require integration over many steps, because they accurately
reflect an underlying iterative physical process, and are thus typically more
difficult to analyze mathematically.

To make these ideas concrete, and draw out the critical role of local
mechanisms for realistic physical models, we consider a few examples:

Newton's theory of gravitation (still widely used) is a calculational tool
that enables gravitational effects to be computed in terms of the respective
masses ($m_1$, $m_2$) and distance $r$ between the centers of mass of two
bodies:
\begin{equation}
  F = G \frac{m_1 m_2}{r^2}
\end{equation}
But this is not a physical model that could function autonomously, because the
math requires one to somehow know the physical distances between relevant
objects (and their respective masses), and not only is this a nonlocal
computation, there are a potentially infinite number of other bodies that need
to be taken into account.  The very notion of a celestial body of
gravitational importance requires expert judgement, and is an example of a
virtual object from the perspective of fundamental physics.  In a physical
model, one would expect that gravitation actually derives from the collective
effects of each individual atom within all the different celestial bodies in
the universe, at which point the Newtonian computation is completely
unworkable and absurd.

Einstein's general theory of relativity, on the other hand, shows how
entirely local, speed-of-light propagation of spacetime curvature, operating
according to uniform functions at each location in space and time, can
convey gravitational forces without any of the problems associated with the
Newtonian calculational tool.  It is a true physical model of the first
order: the mathematical constructs map directly onto physical processes that
are entirely plausible and compelling for what nature can be autonomously
doing to produce the phenomenon of gravitation.

Coulomb's law for the strength of the electric field as a function of
distances between charged particles is very similar to Newton's gravitational
formula, and similarly represents a useful calculational tool, but is not a
good model of how physics actually operates, for all of the same reasons.
Similarly, the Coulomb gauge formulation of Maxwell's equations implies
immediate action at a distance for the electrical potential, which is clearly
incompatible with special relativity.  It turns out that some nonlocalities in
this framework actually cause the observed EM fields to still propagate at the
speed of light, but one can still get into trouble using this gauge
incorrectly \citep{BrillGoodman67,Jackson02,Onoochin02}.

In contrast, Maxwell's equations in the Lorenz gauge provide a very
appealing physical model of electrodynamics, involving simple local wave
propagation dynamics operating on the four-vector potential:
\begin{equation}
  \covgr \congr \con{A} = \cov{k} \con{J}
\end{equation}
where the four-potential is: $\con{A} =(A_0, \vec{A}) = (A_0, A_x, A_y, A_z)$,
and the four-current is: $\con{J} = (\rho, \vec{J}) = (\rho, J_x, J_y, J_z)$,
and the four-constants are: $\cov{k} = \left(\oneo{\epsilon_0}, \mu_0, \mu_0,
  \mu_0 \right)$.  In this physical model, EM waves naturally propagate at the
speed of light, everything is automatically consistent with the constraints of
special relativity, and it is again easy to imagine how autonomous physics can
happen like this.

The clear pattern here is that plausible autonomous physical models leverage
{\em local} propagation of signals according to simple laws, which avoids the
immediate difficulties that are encountered in nonlocal frameworks.  Once
interactions become nonlocal, they typically become infinite in scope, because
anything can potentially influence anything else, and thus any kind of
autonomous physical process becomes inconceivable.  How is nature possibly
going to manage all this infinite bookkeeping?  A great way to appreciate this
difficulty is to attempt to construct an autonomous computer algorithm that
implements a nonlocal interaction, in a way that can apply to large scale
systems with many interacting entities --- the exponential character of these
systems makes them essentially intractable.  And yet nature just does it,
autonomously and ``effortlessly''.

The currently popular gambit that one can leverage the supposedly nonlocal
quantum computations that nature is performing to do more powerful computation
than a regular computer, can be turned on its head: exactly how is nature
doing all this nonlocal computation, really?  Is there any possible way to
efficiently carry out a nonlocal computation, that nature could be leveraging?
If there were, we could just implement that on a regular computer.  Thus, the
very promise of quantum computers suggests that we have absolutely no idea
how nature could possibly function at the quantum level, if it is truly
nonlocal.

In short, it seems that the local nature of physical laws is an essential
component for an autonomous, objective physical reality.  More than any other
factor (e.g., deterministic vs. stochastic behavior) the issue of locality
vs. nonlocality is a fundamental dividing line between physical models and
calculational tools.

It should be clear that both calculational tools and physical models play
essential and complementary roles in the field, and should in no way be
construed as mutually exclusive (even though people inevitably do).  Even
though physical models are often not convenient frameworks for calculation,
they play a crucial role in grounding and constraining physical intuition,
which should then inform the application of calculational tools.  In
particular, calculational tools often contain shortcuts and simplifications
relative to the underlying physical model, and one can obtain nonsensical
results if these are not appreciated (e.g., accidental violations of
speed-of-light propagation in the Coulomb representation).

\section{The Unfortunate State of Quantum Physics}

In quantum mechanics (QM), there are no physical models, only calculational
tools, and most physicists would likely argue that it is impossible to develop
a realistic physical model.  The strict Copenhagen interpretation of QM
specifically disavows the notion of a physical reality outside the scope of
measurements, and Everett's Many Worlds interpretation postulates that the
universe splits into multiple different copies at each measurement event.  We
don't need to belabor how insane this all seems, and yet physicists have
resigned themselves to just accepting this insanity, and look upon those who
protest it with disdain.  The motto appears to be, ``shut up and calculate'',
which is precisely consistent with a field dominated by calculational tools.
But from the arguments above, made on firmer ground, it is clear that one
should harbor a strong mistrust of nonlocal calculational tools for telling
you with perfect accuracy about how a physical system should behave.
Unfortunately, without a plausible physical model, physicists have nowhere to
turn, and the resulting insanity just seems to compound itself, with ever more
bizarre mathematically-motivated ideas being passed off as physical models
(e.g., curled-up invisible extra dimensions, multiverses, etc).  Put on your
tin foil hats folks, and be on the lookout for wormholes to those carefully
hidden extra dimensions and parallel universes!

What if all this insanity is just based on an unfortunate sequence of
misunderstandings, which have hardened over the years into an unnecessary
rejection of the possibility of obtaining a physical model?  That would be
embarrassing!  But I can't quite seem to escape this conclusion.

From the beginning, it seems that those who have advocated for realistic
physical models of quantum mechanics have adopted an overly restrictive set of
constraints for what features a physical model must possess, and much of the
debate has been distracted by the resulting confusion.  In particular, there
has been a strong focus on determinism, for example Einstein's claim that
``God does not play dice..'', and the specific idea that a hidden variable
model should {\em simultaneously} determine the outcomes of all possible
measurements.  But as we can see from the consideration of physical models
above, {\em locality} is by far the more important constraint, and in fact the
strong determinism of Einstein seems entirely inappropriate when considering
the fundamental wave nature of quantum physics as we discuss more later.

Thus, we focus for now squarely on the issue of locality, in the context of
the strongest existing demonstrations of quantum nonlocality.

\section{Quantum Entanglement, Nonlocality, and Bell's Inequalities}

The strongest case for quantum nonlocality comes from John Bell's treatment
\citep{Bell64,Bell87} of a thought experiment initially conceived by
\citet*{EinsteinPodolskyRosen35} (EPR).  There is a massive literature on this
topic, so this discussion will be very brief, although the specific path of
reasoning here seems relatively novel and very clearly illustrates the
absurdity of the standard QM assumptions when applied to ``photons''.  The key
idea is that according to QM, if two particles happen to interact locally in
such a way that their states become {\em entangled}, and they then separate
from each other, their states somehow remain interconnected, such that a
measurement conducted on one particle (call it A) will predict the outcome of
a separate measurement on the other particle (B).  Mathematically, it is said
that the two particles share the {\em same} state function, which is in a
state of uncertainty (a superposition of multiple possible states).  To
greatly simplify the basic predictions of QM in this case, we consider instead
what would happen if you perform two {\em successive} measurements on the {\em
  same} particle, instead of two separate measurements on the two different
entangled particles --- these two situations should be mathematically
identical according to QM.

If A and B are photons created (as they are in the relevant experiments
described later) from an atomic cascade or spontaneous parametric down
conversion, then they should be entangled, and either have the same
polarization or 90 degree opposite polarization (depending on the details of
the procedure).  A polarization measurement is made by placing a photodetector
behind a polarizing filter.  If we happen to know the exact polarization angle
of a light source (call it $\theta_s$), a classical EM result known as Malus's
law states that the polarization filter will allow $cos^2(\theta_s-\theta_f)$
amount of light through, where $\theta_f$ is the angle of the filter.  This is
100\% if the angles are the same, and 0\% if they are 90 degrees apart, and
somewhere in between otherwise.  You can try it out yourself by tilting your
head while looking at your laptop or any other LCD screen with polarized
sunglasses on.

Using these facts, we can calculate what would happen if we could make two
successive polarization measurements on the same photon (the photodetector
absorbs the photon so this technically isn't possible in this case, but it
is with other isomorphic cases).  The result of the first measurement (M1) is
always going to be completely random, because the polarizations of the
photons are unknown and unlikely to be biased in any way, and the angle of
the polarization detector used for M1 is totally arbitrary.  However, at the
next measurement (M2), we can make some very strong predictions.  If M2 is set
to the same angle as M1, then there should be a 100\% {\em coincidence rate}
between the two detectors.  That is, if M1 registers a detection event, then
M2 should as well, and vice-versa.  And if M2 is 90 degrees off of M1, there
should be a 0\% coincidence rate.  And for any angle in between, the
probability of M2 firing given that M1 did should be
$cos^2(\theta_2-\theta_1)$ based on the respective angles of the detectors.

Here is the first critical point: {\em these predictions are true regardless
  of how you rotate the polarizing detectors}.  The only way for this to be
the case is if {\em the first measurement actually rotates the polarization of
  the light to align with its filter.}  Otherwise, if instead you thought that
the photon had some specific polarization angle that remained unchanged by
M1, the results of M2 would be given by $cos^2(\theta_s-\theta_2)$, where
$\theta_s$ is this true ``source'' polarization angle.  This case would
clearly have absolutely no relationship to the angle on M1.  This difference
of a cosine relationship between the two detector angles vs. an
independent relationship between the two detector angles is the basis for
Bell's inequalities, which simply quantify this difference in a way that is
amenable for empirical tests.  However, in this case of two sequential
measurements, it is trivial to conduct this experiment yourself and see the
results.  Just take two polarized sunglasses and rotate them relative to each
other.  It doesn't matter what angles you choose, you'll always observe that
the first polarizer does indeed rotate the light to its angle, so that there
is always a $cos^2(\theta_2-\theta_1)$ function to the light intensity that
makes it through both glasses.  Note that we don't actually measure the light
after the first polarization step, so we avoid that problem in this
experiment. 

From this experiment, it is obvious that the ``measurement process'', at least
for light waves, does not immaculately reveal the ``true'' polarization state,
but rather reflects an interaction between the incoming light wave properties
and the properties of the measuring device.  The measuring device imposes a
good bit of its own ``reality'' onto the state of the light wave.  In QM
terminology, this means that the measurement is {\em contextual}
\citep{Shimony84,Gudder70,Khrennikov01,Rovelli96}.  I cannot imagine any
measurement taking place on a wave that would {\em not} be contextual in this
way.  It also appears to be true of spin measurements on electrons.  Waves are
way too fluid a thing to stand up to the kind of abuse applied by a
measurement device unscathed!  Einstein apparently didn't seem to catch the
wave vibe from quantum mechanics, and postulated that measurements should {\em
  not} be contextual, and instead should reflect some kind of deeper hidden
variables possessed by particles.  But this doesn't even explain the behavior
of classical EM for successive measurements, as we just saw, and seems
completely untenable.  It would seem that only a strong adherence to the
particle model (as Einstein unfortunately maintained), with the hard little
particle somehow possessing very definitive properties, would motivate such a
belief.

And now for the second critical part: We take our results from the two
sequential measurements, and attempt to apply them, as QM says we should,
directly to the two separate ``photons,'' A and B.  The theory says (with a
straight face), that if we perform polarization measurement M1 with a given
angle on photon A, it must somehow influence things such that measurement
M2 on B obeys the very same equation as for two sequential measurements:
$cos^2(\theta_2-\theta_1)$.  Yes, the angle of the M1 polarizer must somehow
influence the behavior of the measurement process on B.  Even if A and B have
had enough time to fly {\em arbitrarily} far apart (e.g., in principle to the
opposite ends of the universe).  It gets better: this is all supposed to
happen absolutely instantly.  No time delay at all.

To which I say: ``Surely you must all be joking!?''  This is a good candidate
for the most fantastical, absurd prediction in the history of science, and
{\em nearly everyone in quantum physics swallows it whole.}  It is a
completely non-physical, non-local, non sequitur.  The physics of two
sequential measurements on one photon versus two separate measurements on
two separate photons are entirely different, and it is just not clear why
anyone would think they should correspond to the exact same thing.  This seems
like a classic case of the calculational tools of QM being misapplied, and a
cross-check with some kind of physical model would quickly reveal the error.
But because there is no accepted physical model in QM, there really isn't a
suitable fallback position, and so people just seem to accept what the
calculational tools tell them.

However, there really {\em is} a de-facto widely accepted physical model for
the behavior of photons, and it is none other than classical EM (i.e.,
Maxwell's equations).  As we discuss in greater detail later, the
semiclassical approach is widely used to calculate a great variety of quantum
optics effects.  If we apply the classical EM physical model to the two photon
entanglement case, it is patently obvious that no such entanglement phenomena
should or could be observed.  As we just described above, the physics behind
the two successive measurements is completely obvious and sensible --- you can
visualize the polarization of the light waves rotating around as they pass
through the M1 filter, thus affecting the results for the M2 measurement.

The extension to two separate photons makes absolutely no sense --- how can
a local interaction between a polarization filter and a light wave possibly
affect a similar such interaction separated by an arbitrary distance, when the
two light waves have been traveling apart at the speed of light!?  There is
simply no way within the classical EM framework for light waves to continue to
interact once they start heading in different directions --- everybody's moving
at the speed of light, and nothing sticks around in between to mediate any
kind of connection between them.  Furthermore, EM waves do not even have any
way of interacting with each other --- there is no physical basis for any kind
of ``signal'' to be sent from one EM wave to another --- the only medium for
such a signal would be the EM field itself, and it just passes right through
due to linear superposition.  Thus, there is no plausible mechanism that could
mediate an entanglement state in the first place, at least according to the
classical EM model.

To understand how the field could have come to this point of unquestioning
belief in the applicability of entanglement to light, we next review the
shocking state of the QM approach to light, which reveals how stunningly
non-physical a calculational tool the standard QM model (QED) is.  Indeed,
practicing physicists apparently avoid using QED whenever possible, and
instead rely on good-old-fashioned classical EM in the context of the
semiclassical approach, which is reviewed thereafter.

\section{The Quantum Theory of Light}

The notion of a photon conveyed by descriptions of paradigmatic quantum
phenomena such as the photoelectric effect described by Einstein in 1905
strongly connotes a physical entity that is absorbed and emitted by atoms as
they jump up and down in energy levels \citep{Klassen11}.  The photon has a
frequency, and we inevitably picture some kind of little wave packet vibrating
with that frequency, somehow combined in an inexplicably weird way with a hard
little photon particle.  For all the myriad discussion of photon-based
phenomena in QM textbooks (e.g., the two slit experiment), one naturally
assumes that the Schr\"odinger wave packets shown in the diagrams actually
represent these photon wave packets.  However, I was astounded to discover
that until relatively recently \citep{BialynickiBirula94,Sipe95}, photons had no
proper Schr\"odinger-like wave function in the standard QM framework.
Instead, the only actual mathematical treatment of the photon concept comes
from Quantum Electrodynamics (QED), which has such a strange representation of
a photon that is essentially unrecognizable from the above picture (see
the excellent papers in the 2003 special issue of {\em Optics and Photonics
  News Trends} \citep{RoychoudhuriRoy03} for expert discussion, and e.g.,
\citealt{Lamb95}). 

QED is a calculational tool {\em par excellance}, in part because it is based
on a Fourier transformed representation, replacing the usual space and time
coordinates with frequency and phase coordinates.  Mathematically, Fourier
space can represent any kind of wave function that might occur in the real
world, and lots of things are more convenient to calculate in Fourier space.
But I was truly shocked to finally understand that the QED notion of a photon
is actually defined in this Fourier space representation, instead of in the
real spacetime coordinates: a photon is a fundamental mode of vibration at a
given frequency in Fourier space. As such, a photon is {\em intrinsically
  nonlocal}, spanning across the universe in effect!  To obtain any kind of
localized wave function, you need to combine many different photons at
different phases and frequencies.  Thus, the perfectly intuitive wave packet
model of the photon sketched above in fact corresponds to a huge raft of QED
``photons,'' constructed just so as to produce what one would otherwise assume
occurs quite naturally through the emission and absorption process.

This Fourier space is quantized, in a process referred to as {\em second
  quantization}, making it a Quantum Field Theory (QFT), known as a Fock
space.  This allows you to count the number of different photons associated
with each vibrational mode, and there are operators that create and destroy
these photons.  This ability to formalize the creation and destruction process
appears to be one of the great advantages of the QED framework, making it such
a useful calculational tool.  But the cost of this move is being stuck in this
nonlocal, nonphysical Fourier space: this quantum field is nothing like a
classical EM field.


This second quantization process apparently leads directly to the notion of a
zero point field (ZPF), which is the ground state of the quantum field.  The
ZPF describes the quantum state of the vacuum, and a huge amount of the power
of QED comes from the ``vacuum fluctuations'' --- energy can be temporarily
``borrowed'' from the vacuum to create ``virtual'' particles which then
interact with ``real'' particles, in ways that produce subtle but measurable
effects.  It is also said that these virtual particles mediate the actual EM
forces, e.g., by electrons passing them back and forth all the time.  In
contrast, the classical EM equations (e.g., in the Lorenz gauge shown above)
explain the same EM forces without any reference to these mythical ``virtual''
entities.  And classical EM does so in a very local, emergent, physically
compelling way, in contrast to this clumsy picture of electrons somehow
tossing virtual balls back and forth to each other.  Discretizing a continuous
force field into imaginary virtual particles definitely seems like a major
step backward.

In any case, it is evident that the constructs in the calculational tool of
QED can be mapped onto other kinds of constructs in other frameworks.  And
when one framework has to label a majority of what it contributes with the
term ``virtual'', it seems obvious that it is a tool, not a physical model.
Unfortunately, they didn't label the entire apparatus of QED with the term
``virtual'' --- the fact that some things were explicitly virtual somehow
tends to give the other non-virtual things more credibility, but really the
notion of a photon in this framework is just as virtual as anything else.

The math of QED is based on infinite sums, and these quickly diverge to
infinity.  Through heroic efforts, a scheme for renormalizing away these
infinities was constructed, and the resulting framework is rightly celebrated
for being able to calculate extremely accurate numerical predictions, which
fit all the known experimental results.  Thus, it is clear that this tool
captures something very powerful and true about the way physics works, but any
attempt to ascribe physical reality to its central constructs, e.g., the
notion of a photon, or passing virtual particles around, seems like pure
folly.  Indeed, several of the developers of this framework regarded it with
disgust and disappointment, even as they used it to compute important results
\citep{Jaynes78,Grandy91}.

In summary, the QED calculational tool is intrinsically nonlocal and its central
notion of a photon is manifestly non-physical.  Thus, it is perhaps not
too surprising that the crazy non-local behavior ascribed to photons in
the entanglement case doesn't seem that objectionable to most physicists.  But
the recently-developed ``first quantized'' wave function for photons, and
the classical EM framework, both seem at odds with the QM entanglement
prediction as described above.  We discuss next that there is no solid basis
in any existing experimental data to reject these localist models as plausible
physical models underlying the phenomena that QED describes in its own weird
way.

\section{Semiclassical Models}

Another shock in digging deeper into the literature was that the vaunted
photoelectric effect, the paradigmatic quantum phenomenon that gave birth to
the notion of a photon, can be accounted for within the semiclassical or
neoclassical approach, where the EM field is treated entirely classically
(i.e., Maxwell's equations), and only the atomic system is quantized
\citep[e.g.,][]{JaynesCummings63,Jaynes73,Mandel76,Barut91,Grandy91,GerryKnight05,MarshallSantos97}.
Not only the photoelectric effect, but literally every other major phenomenon
that was initially thought to uniquely reflect the existence of a photon
particle (or more accurately, the formalism of QED and second quantization of
the EM field) has been accounted for within the semiclassical approach,
including the Lamb shift, Compton scattering, vacuum polarization, the
anomalous magnetic moment of the electron, etc.

The frontier is ever expanding, however, and the most recent debates concern
various photon statistics effects, for example an anticorrelation effect
described by \citep{GraingerRogerAspect86}, and the antibunching effect
described by \citep{HongOuMandel87}.  These results can be accounted for with
a version of semiclassical theory that also contains the zero point field
(ZPF) from QED, which go by the name stochastic electrodynamics (SED)
\citep{MarshallSantos88,MarshallSantos97}.  It is also possible to account for
some of the effects due to limitations of the apparatus
\citep{Sulcs03,SulcsOsborne02}.  Pushing back the other way,
\citet{Marshall02} showed that the SED model of spontaneous parametric down
conversion also predicts spontaneous parametric {\em up} conversion, which
apparently is not something that QM would predict.  Recent evidence confirms
this prediction \cite{SunZhangJiaEtAl09,AkbarAliEtAl10}, which both supports
the SED model and challenges the conventional QM model.

To provide some context for the role of the ZPF in the semiclassical approach,
it is important to appreciate that the semiclassical approach is based on
approximations to the fully coupled Maxwell-Dirac system, which we discuss in
greater detail later as a plausible physical model.  These equations are
relatively simple to analyze individually, but very complex when coupled in
the natural way.  Importantly, the Maxwell-Dirac coupling naturally produces a
{\em self-field} or {\em radiative reaction} produced by the electron back
onto itself.  Many of the phenomena accounted for through the ZPF in QED can
be traced to the self-field in the Maxwell-Dirac system
\citep{Milonni84,Barut91,Grandy91}.  As \citet{Jaynes78} nicely explains,
the only vacuum fluctuations in the ZPF that really matter are the ones right
near the electron, and in fact the electron is directly responsible for
creating these fluctuations in the first place through its self-field.  It
remains unclear (to me at least) if this argument goes all the way through for
the photon statistics effects described by the SED theory, and there appears
to be some ambiguity about whether everything can truly be accounted for just
by the self-field \citep{Milonni84,Grandy91}.

Despite all these successes in terms of seemingly compelling physical models
of otherwise mysterious quantum phenomena, it seems that most people regard
semiclassical approaches as being fundamentally wrong, and thus not worth
changing their world view over.  No single semiclassical model can account for
all the relevant data, and there are various problems found in each of the
different models \citep[e.g.,][]{Mandel76}.  For example, there is apparently
a {\em chirp} (corresponding to a dynamical transition of some sort) in the
Lamb shift of Jaynes' model, which is not present in QED and has not been
found experimentally.  However, other semiclassical models do not make this
prediction apparently \citep{Milonni84}.  Another problem concerns the rate of
absorption of a quantum of energy from the EM field into an atomic system ---
can it happen as quickly as the instantaneous effects predicted by QED, which
seem to be consistent with experimental data \citep{Mandel76}?  Interestingly,
\citet{Mandel76} shows that this problem goes away with a wave packet
model of the photon, which seems quite plausible.  He however then goes on to
argue that such a model is inconsistent with the photon anticorrelation data,
but SED shows that this may not be a problem.

The bottom line appears to be that to convince a physicist of something, you
have to be able to derive accurate calculational results, and short of that,
the physical plausibility of the model does not count for much.  Even if there
is every indication that the full Maxwell-Dirac system could potentially
account for all the relevant results, and at least no solid evidence that it
cannot, the fact that it is too difficult to analyze renders it largely
irrelevant in the daily life of a practicing physicist.  And all of the
various approximations that are more analytically tractable and form the basis
of the semiclassical approach have clear limitations, so they are not worth
bothering with either, unless they turn out to be useful calculational tools
within their realm of applicability (which actually appears to be the case in
a number of instances, e.g., \citealt{GerryKnight05,RoychoudhuriRoy03}).

But from the outside looking in, and putting an absolute premium on finding a
local physical model that could potentially explain the quantum world, the
semiclassical approach seems exceptionally promising.  Despite every attempt
to prove otherwise, the classical Maxwell's equations provide an incredibly
accurate, and appealing, physical model.  In every case, the origins of
quantum weirdness can be traced back to the behavior of atomic systems (and
perhaps the ZPF), not to the EM field itself.  At the very least, it would
seem difficult for someone to strongly refute this possibility.  Thus, by
extension, the assumption that light should somehow obey the strange quantum
property of entanglement seems entirely suspect, and given how much weight
this carries for our fundamental understanding of the nature of physics, it
seems nothing other than completely insane that the entanglement of light goes
apparently unquestioned by the vast majority of physicists.

A major factor for this state of affairs is the seeming confirmation of the QM
photon entanglement predictions in a series of experiments testing Bell's
inequalities, as we discuss next.



\section{Experimental Tests of Bell-type Inequalities}

A number of experiments using entangled photon sources with separate
measurements of polarization (as described earlier) have been conducted, and
their results appear to confirm the QM entanglement predictions
\citep[e.g.,][]{AspectDalibardRoger82,AspectGraingerRoger82,TittelEtAl98}.  In
one case, the two measurements were separated by 10km \citep{TittelEtAl98}!
Although discussion of these experiments is dutifully accompanied with mention
of certain ``loopholes,'' these are very often presented as contortionist
exercises in deflating the orthodox interpretation.  People complain that one
loophole is used in one case, and another in another case --- does nature
choose which loophole to exploit depending on the circumstances?

In delving deeper into this literature, I was again shocked to find that these
so-called ``loopholes'' appear instead to be accurate physical models of the
actual experiments, and there are very good physical reasons to consider
different such ``loopholes'' for different experimental situations.  The major
``loophole'' for the experiments based on photons is known as the
detection/fair sampling loophole, which basically states that the QM
predictions depend on the detectors reporting a fair sample of the photons
that are generated from the source, and enough of them to make sure that all
the relevant statistics are being counted.  Well, it turns out that even the
best current photodetectors can only detect up to 30\% of the photons, and
furthermore, there are strong physical reasons to believe that the
polarization angle strongly influences the detection probability, violating
the fair sampling assumption.  Detailed models of this sort can reproduce the
observed data quite accurately, for a variety of experimental configurations
\citep[e.g.,][]{MarshallSantosSelleri83,MarshallSantos85,Thompson96,AdenierKhrennikov03,Santos05,AschwandenEtAl06,AdenierKhrennikov07}.
Interestingly, one of these analyses \citep{AdenierKhrennikov07} shows that
accepting the fair sampling assumption produces results that violate the ``no
signalling'' property of the standard QM prediction, strongly implicating that
fair sampling has been violated.

As for the other major loophole, amusingly enough called the ``locality''
loophole, it pertains to experiments on massive particles, which are
apparently the only ones that can practically close the detection loophole
(with rates $> 90\%$; \citealt{RoweEtAl01}).  If locality is considered a
loophole, something is seriously wrong with the term ``loophole''.  And the
distinction between massive and massless (photons) that determines which
``loophole'' applies is anything but arbitrary.  Two massive entangled
particles can always communicate via light-speed interactions (e.g., EM waves)
by virtue of the Lorentz contraction effects of special relativity, which
ensure that even when massive objects are moving near the speed of light,
light still moves at the speed of light relative to them.  Indeed, in the
\citet{RoweEtAl01} experiment, the two atoms in question were strongly
interacting via a Coulomb (EM) force, over a very short distance.
Furthermore, there are other problems associated with these experiments
related to errors in the measurement angles \citep{Santos09}.

Thus, again, one cannot help but conclude that any reasonable person who
appreciated the true importance of the construct of locality for understanding
how nature actually works, would recognize that these experiments provide woefully
ambiguous support for the standard QM model of entanglement, and indeed could
be seen as providing increasingly strong support {\em against} the standard
view, given the increasing passage of time without a more definitive
experiment that overcomes the ``loopholes'' \citep{Santos05}.

In the next section, we revisit the assumptions that lead to the QM
description of entanglement, and consider how the calculational tool of QM may
prevent a more accurate description of the underlying physical processes in
terms of the distinction between massive and massless particles.

\section{Quantum Entanglement Revisited}

Why does QM predict this bizarre entanglement phenomenon in the first place,
and is there some way to generalize the theory that would accommodate a strong
locality constraint on entanglement?  These are generally questions beyond my
ability to answer, but here are a few physical intuitions that may be of
relevance --- they certainly help me feel like I understand better what is
going on.

The central features of entanglement that need to be captured in any framework
are that the states of the two particles are unknown, and yet there are
strong constraints on their states (either they must be the same or opposite,
depending on the specific case in question).  Representing exactly this kind
of situation {\em for a single particle} is the fort\'{e} of QM: the unknown
state is represented by a superposition of multiple possible states, and the
strong constraints come from the basic conservation laws built into QM, which
define how states are affected by measurements.  Interestingly, the
mathematics of QM can be derived very generally from certain kinds of
conservation laws, suggesting that the standard QM formalism is really just an
abstract probability calculus, with these strong conservation laws, which
manifest as a requirement for continuous reversibility \citep{Hardy01}, or
in the purification postulate \citep{ChiribellaDArianoPerinotti11}.

One mental image for this is that the measurement process in QM is really just
about {\em rotating} things around in state space, e.g., on the surface of a
(Bloch) sphere --- you never lose (or gain?) any information about the system
in question, you just rotate that information around on different axes.  If we
go back to our polarization detectors, these just rotate the polarization
state of the photon around to different angles, but do not fundamentally
alter the magnitude of the polarization property itself.  In contrast, if the
measurement process did not rotate the polarization state of the photon,
then it would be possible to setup a sequence of measurements that eliminate
the polarization state entirely --- it would end up with no measurable
polarization at all!  Hence, the contextuality of the measurement process is
really just a manifestation of this conservation principle that lies at the
heart of QM.  Another potentially useful image is a ball of mercury --- you
can squeeze it into many different shapes, but it fundamentally conserves its
overall properties.  If you try to measure how tall it is, that squeezing
process may cause it to squirt out in the horizontal dimension, and
vice-versa.  This captures the fundamental uncertainty principle --- squeezing
things one way causes them to squirt out in other ways, meaning that you can't
measure both properties simultaneously.

All of this makes sense for the state of a single coherent entity (a
``particle''), which is generally indivisible and really should always behave
like that tight little ball of mercury.  But does it make sense for two
separate entangled particles?  Mathematically, QM represents the two entangled
particles just like a single unknown particle, because that is presumably the
only way to capture the appropriate properties of the state being in a
superposition and yet strongly constrained.  But what if QM could be extended
to represent nonlocal entanglement in a different way?

For example, one critical question about the QM conservation principle is,
over how big of a system do you need to apply it?  When you make a
measurement, we know that the measuring device imparts its state onto the
state of the ``particle.''  But wouldn't it also make sense that the
particle imposes some of its state onto the measuring device?  Under this
view, what is conserved is the total state of the particle + measuring
device, not just the particle by itself.  Does this have any implications
for the entangled case?  It might suggest that we treat M1 + A as one quantum
state, and M2 + B as another, separate quantum state, each of which then will
obey the proper conservation dynamic.  But this will lose the constraint that
A and B share some critical property, which nevertheless remains unknown (in a
state of superposition for both).  So really you need to represent the whole
thing: A + B + M1 + M2 as one big state.  But this is then a manifestly
nonlocal state.  Nevertheless, this is how it is routinely done in QM --- many
QM states are defined in high-dimensional, nonlocal configuration space.

There are a few observations we can draw from this:
\begin{itemize}
\item The calculational tools of QM are very much like a Coulomb or Newtonian
  representation of the problem --- they strongly encourage nonlocal
  configuration space representations, and really all the physics is being
  captured by virtue of this fundamental rotational conservation property of
  the quantum world, which can be applied at many different levels of
  analysis.  As we know from the EM and gravitational domains, the existence
  of a convenient nonlocal configuration-space calculational tool does {\em
    not} preclude the existence of a local realistic physical model.

\item There does not seem to be any way in the QM calculational tool to
  represent the presence of two unknown states that are nevertheless
  constrained to be initially identical (or opposite).  This raises the
  possibility that entanglement is a kind of mathematical accident of the
  limitations of the calculational framework --- it just cannot represent this
  state accurately.  Note that there does not need to be any kind of violation
  of the all-important conservation laws for this case, because M1+A and M2+B
  can each still be conserved, and A and B each just rotate around anyway ---
  they just have some kind of shared heritage that cannot be properly
  represented.  This could be construed as a very weak form of hidden
  variables.
\end{itemize}

These considerations suggest that an underlying physical model, which obeys
the locality constraint, might give rise to a spectrum of entanglement
scenarios.  In the most obviously entangled case, you have particles that
{\em remain in close physical proximity} and are thus {\em continuously
  entangled} --- it seems clear here that a first measurement M1 on
particle A would likely produce strong disruption of the state of
particle B, such that a second measurement M1 on B would very plausibly be
affected by M1, exactly as the standard QM entanglement model holds.  In this
case, the underlying physical model accords well with the assumptions required
from the calculational tool, and everything is consistent. The case of
entangled photons represents the other extreme, which could be described
as {\em formerly entangled}, and is simply not representable within the QM
formalism.  Hence all the confusion surrounding this erroneous case.  In
between, one might imagine some kind of continuum, where some degree of
continued interaction produces some level of correlation in the measurements,
but not as strong as one would expect from the continuously entangled case.

This spectrum is based on the idea that physical locality drives entanglement,
which seems to be an important component of the standard QM model already.
Specifically, the {\em source} of entanglement in the first place is directly
tied to physical proximity according to the conventional description.
Therefore, it doesn't seem to be a particularly radical suggestion that the
continued maintenance of entanglement should also depend on continued physical
proximity.  It is not clear how to mathematically integrate this locality
constraint into the QM formalism, but given that it is merely a calculational
tool, it is to be expected that there are things that it cannot accommodate.

Lastly, we consider the actual paradox behind the EPR thought
experiment \citep{EinsteinPodolskyRosen35}.  This paradox is that the
entanglement scenario appears to allow one to determine more information than
would otherwise seem possible about the state of a particle, by performing
separate measurements on each of two entangled particles, instead of two
sequential measurements on a single ``particle.''  As initially formulated,
this paradox was erroneous from the standard QM perspective, because EPR
assumed that the two measurements would not affect each other, and yet that M1
on A would nevertheless tell you something precisely about B.  This is having
your quantum cake and eating it too --- the only way M1 can tell you something
definitive about B is if it actually affects B in exactly the same way it
affects A.  Thus, once M1 affects B and thus M2, then it really is identical
to two sequential measurements, and there is no paradox.

Conveniently, the spectrum outlined above does nothing to introduce a new
paradox.  Never do we adopt the untenable assumption of hidden states that
simultaneously determine all measurements --- each measurement is an
interaction (i.e., contextuality).  For the formerly entangled case of
photons, the outcome of M1 on A doesn't tell you very much about what is
going to happen with M2 on B --- in the case of polarization you really only
know that A (and thus B) is not polarized 90 degrees relative to the angle on
M1 --- it could be 89 or 91 or any other polarity (and this assumes perfect
polarizers which is never possible in reality).  Thus, the ``heritage''
information is much weaker than the continuously entangled case, and much
weaker than what was envisioned in the EPR hidden variables.

\section{Toward a Realistic Physical Model of the Quantum World}

Finally, we conclude with some considerations for what a realistic, local,
physical model of the quantum world should look like.  As noted earlier, the
clearest indication of a calculational tool is the appearance of ``virtual''
entities, which abound in QED for example, and the clearest indication of a
true physical model is the {\em complete absence} of such virtual entities.
In the standard QM formalism, the most glaring virtual entity is the wave
function itself, which is not regarded as physical, and instead only provides
the probabilities for experimental outcomes.

Given that the quantum wave function determines actual physical behavior, it
seems that we simply cannot escape the conclusion that {\em a realistic
  physical model {\em must} have a physically real wave function}.  Together
with the local propagation constraint, this strongly suggests that our model
of the electron should be like that initially suggested by Schr\"odinger,
where the wave function defines the evolution of a distributed mass of charge
density, and that is all there is.  Once you have a real wave function, there
is really no room for the particle concept, which after all is the source of
so many conceptual difficulties anyway (as enumerated below, including the
case of the {\em pilot wave} deBroglie-Bohm alternative model).

Schr\"odinger apparently abandoned this pure-wave model when he realized that
his wave functions had to be defined within high-dimensional configuration
space, and he also thought the spread of the wave packet seemed unrealistic,
given for example the tracks of particles through cloud chambers.  We discuss
these and other concerns after describing the Maxwell-Dirac system as the
preemptively obvious candidate for a realistic physical model that satisfies
the constraints above.

\subsection{The Maxwell-Dirac System}

The two primary wave fields of interest are the classical EM field coupled
with a wave function describing the behavior of the electron (other such
fields concern the strong nuclear force presumably, but we can focus on the
basic EM phenomena to explore the relevant issues).  As reviewed above, the
best physical model of the EM field is Maxwell's equations in the Lorenz
gauge:
\begin{equation}
  \covgr \congr \con{A} = \cov{k} \con{J}
  \label{eq.maxwell}
\end{equation}
The most accurate physical model for the electron, which is manifestly
covariant and thus automatically compatible with special relativity, and
describes a wave of charge density, is given by Dirac's equation, which can be
written (unconventionally) in a second-order form that strongly resembles
Maxwell's equations:
\begin{equation}
  \left[\left(i \hbar \covgr - \frac{e}{c}\cov{A} \right)^2
      + \frac{e}{c} \vec{\sigma} \cdot \left(\vec{B} + i \vec{E} \right) \right] \dw 
     = m_0^2 c^2 \dw
  \label{eq.dirac}
\end{equation}
where $\dw$ is a state field defined over standard 3 dimensional space that
has two complex numbers (i.e., 4 degrees of freedom, coincidentally the same
number of degrees of freedom in the four-potential $A$), $\vec{E}$ and
$\vec{B}$ are the electric and magnetic fields derived directly from $A$, and
$\vec{\sigma}$ are the standard Pauli matricies that describe the electron's
spin:
\begin{equation}
  \vec{\sigma} = \left(\mattwo{0}{1}{1}{0}, \mattwo{0}{-i}{i}{0},
    \mattwo{1}{0}{0}{-1} \right)
  \label{eq.paulis}
\end{equation}
The charge density $\rho$ and current density $\vec{J}$ of this Dirac electron
are given as follows, which then constitute the four-current $\con{J} = (\rho,
\vec{J}) = (\rho, J_x, J_y, J_z)$ that is the driving source in Maxwell's
equations:
\begin{equation}
  \rho = \frac{i \hbar e}{2m_0c^2} \left(\dw^* \pdt{\dw} - \dw \pdt{\dw^*} \right)
\end{equation}
\begin{equation}
  \vec{J} = - \frac{i \hbar e}{2m_0} \left(\dw^* \vgrad \dw - \dw \vgrad \dw^* \right)
\end{equation}
Critically, charge is conserved by the Dirac equation.

A great deal of confusion has surrounded the fact that the energy associated
with the Dirac wave can be negative.  Although Dirac initially came up with a
somewhat contrived solution to this issue involving a sea of electrons filling
the vacuum states, it is now understood that the Dirac equation describes both
the electron and its antiparticle, the positron, with the positron
corresponding to the negative energy solutions.  There is an intimate
relationship between the electron and positron, for example the fact that they
can be created from just high energy EM waves, and similarly when they collide
they annihilate back into high energy EM radiation.  The Dirac equation
naturally handles this most mind-bending of phenomena.

This fully coupled Maxwell-Dirac system could in principle provide a fully
accurate representation of quantum electrodynamics, with only two main
equations and entirely local, simple propagation dynamics.  Unlike so many
attempts to understand what the electron spin is in terms of a hard little
particle somehow spinning, we can instead just take the 4 dimensional state
values in the $\dw$ electron wave function as a literal substrate over which
the Dirac wave dynamics operate, in exactly the same way the 4 dimensional
potential $\con{A}$ is the substrate on which Maxwell's wave equations
operate.  In the cellular automaton framework, these state values are just
that: local state values that the local equations update as a function of
their neighbors.  In this case, spin just amounts to the fact that the state
values are constantly rotating through each other, corresponding to the {\em
  zitterbewegung} (``trembling motion'') of the electron.  It is a
parsimonious, appealing physical model.  I find it astounding that this simple
set of coupled equations could possibly describe much of what happens in the
universe.  This is exactly the kind of fundamental simplicity you would expect
from fundamental physics.

However, one of the biggest barriers to the exploration of this Maxwell-Dirac
system is that, despite the apparently simplicity of the basic equations, its
aggregate behavior is exceptionally complex to analyze mathematically, having
resisted many attempts to understand its full complexity.  Nevertheless,
considerable progress has been made using simplified subsets of the full
system.  Approximations of this system are the basis for the semiclassical
approach described earlier
\citep{JaynesCummings63,Jaynes73,Barut91,Grandy91,MarshallSantos97}.  The
Schr\"odinger wave equation itself can be derived as a non-relativistic,
first-order version of this second-order Dirac equation.  From these windows
into the full system, we gain considerable insight and confidence that it
could actually describe our physical reality.  Also, it would seem that
numerical simulation techniques could be productively brought to bear on this
system --- it does not appear that this approach has been explored to any
significant degree yet.

If the coupled Maxwell-Dirac system is an accurate physical model, it must
correspond in some way to QED, which we know provides highly accurate
calculational results.  As is nicely explained by \citet{Grandy91} (and
summarized briefly earlier), there are strong reasons to believe that there is
a direct correspondence between the two frameworks, which basically provide
two different mathematical representations for the same underlying physical
process of the {\em self-field} of the electron (also known as the {\em
  radiation reaction}).  This self-field is directly physically manifest by
the coupling of the Maxwell-Dirac equations through $A$ and $J$, whereas in
QED they emerge through virtual particle interactions via the zero point field
(ZPF).  Thus, again we see that the physical model has a simple physical basis
for this effect, while the calculational tool invents something virtual to
account for it.  Regardless of this difference, both frameworks capture the
same key physical phenomena, including the Lamb shift in the spectrum of
Hydrogen emissions, the anomalous magnetic moment, and spontaneous emission
from excited atomic states in terms of the effects of the self-field
\citep[see][for reviews]{Barut91,Grandy91,Jaynes78}.  In addition, other
effects associated with the ZPF in QED, including the Casimir effect, vacuum
polarization, and the Unruh effect can be shown to emerge from the
Maxwell-Dirac self field in the approach taken by \citet{Barut91}.

Interestingly, the notion of a self-field is fundamentally incompatible with
the idea of a point particle, because the self-field becomes infinite at this
point.  For this reason, it is completely neglected within the standard (first
quantized) QM framework, which thus remains incapable of accounting for the
above effects.  In the second-quantized QED framework, the infinity is {\em
  renormalized} away, using a mathematical slight of hand that is apparently
quite complex and only can work for some cases, one of which happens to be
QED.  Renormalization appears to be generally regarded with disgust in the
physics community, but because it works, it is also widely accepted.  It is
however yet another indicator of this strong dichotomy between QED as a
calculational tool with all manner of physically absurd properties, and the
simple elegance and natural physicality of the Maxwell-Dirac system.

A remaining question is whether there are physical phenomena associated with
the ZPF that cannot otherwise be attributed to the self-field effects
\citep{Milonni84,Grandy91}, for example, the photon antibunching effects
described by the stochastic electrodynamics (SED) models
\citep{MarshallSantos88,MarshallSantos97}?  Given the simplifications present
in the existing semiclassical models, it seems at least possible that the more
complete electron wave function provided by the Dirac equation, which includes
for example the high-frequency {\em zitterbewegung} oscillation property,
could produce a self-field that has the same effects as those captured in the
SED models through the ZPF mechanism.  If not, perhaps the model needs to be
augmented with a full-fledged ZPF mechanism, as in SED, but this apparently
comes at a considerable cost due to the physically implausible properties of
the ZPF.  Clearly more work needs to be done here to figure out these and many
other outstanding problems.

\subsection{Can we Really Dispense Entirely with Particles?}

The obvious difficulty in adopting a pure-wave alternative is accounting for
all the phenomena otherwise attributed to particles.  In the standard QM
framework, almost all particle-like effects are associated directly with the
measurement process and an associated collapse of the wave function, which is
one of the most contentious and mysterious aspects of standard QM.  During
this measurement process, the wave function is said to collapse into a
discrete state, which is somehow associated with the concept of the particle.
How can this happen within the pure waves-only model?  In addition to these
problems, we have Schr\"odinger's two concerns mentioned above
(high-dimensional configuration space and apparent localization in the cloud
chamber).  These issues appear to stem from the linearity of the Schr\"odinger
equation, as contrasted with the nonlinear Maxwell-Dirac system.

\subsubsection{Linear vs. Nonlinear Systems and Configuration Space}

The linearity of the Schr\"odinger wave equation means that {\em these waves
  cannot represent any interactions} --- two waves just superpose right
through each other.  Thus, any kind of interaction dynamics between multiple
particles requires a higher-dimensional configuration space.  Physically,
higher-dimensional configuration space doesn't make any sense because the
number of particles is not a constant, so the dimensionality of the space is
undefined (dealing with this fluidity in particle number is one of the major
strengths of the Fock space in QED).  This is a clear indication that
configuration space is a calculational tool, not a physical model.

In contrast, the nonlinearities of the Maxwell-Dirac system mean that it does
{\em not} in principle require high-dimensional configuration space to deal
with multi-particle interactions.  Thus, it is possible that a simple 3+1
dimensional wave state space can represent any number of particles and their
relevant interactions, which is an absolute requirement for a local physical
model.  Fortunately, a plausible basis for thinking that the 3+1 dimensional
Maxwell-Dirac system can describe arbitrary numbers of particles is provided
by \citet{Dorling87}.  He leverages the fact that the Dirac equation
actually describes both electrons and positrons, and fully allows for them to
be created and destroyed.  Thus, the Dirac equation is already a 2 particle
equation with effective creation and annihilation operators, and one can adopt
an argument due to Feynman to generalize this into an N particle equation in a
seemingly mathematically sound manner.

Another strong basis for optimism about being able to remain in real 3+1
dimensional spacetime comes from the density functional theory (DFT), which
represents electrons in atoms in terms of an electron cloud density
surrounding the nucleus, in simple 3 dimensional space
\citep[e.g.,][]{ArgamanMakov00}.  There are various corrective terms that must
be added to account for electron-electron interactions, but overall highly
accurate predictions can be obtained from this system, and it has apparently
become the dominant formalism for quantum chemistry.

More generally, the behavior of electrons in the atomic context seems to be
well-described by wave dynamics.  Contrary to the na\"ive atomic models based
on electrons actually orbiting the nucleus, they instead behave like standing
waves, with no orbital momentum in most cases.  And these standing waves are
somehow superimposed on top of each other in a completely intermingled
fashion.  As captured in the DFT models, the atom really does seem to have a
dense cloud of electron charge surrounding it, and it stretches the
imagination to think of hard little particles bouncing around in this context,
each tied to their own wave functions which are nevertheless completely
intermingled.  

The critical property that the Maxwell-Dirac system must exhibit to accurately
capture atomic behavior in real 3+1 dimensional space is the Pauli exclusion
principle.  Interestingly, although the spin-statistics theorem that underlies
this principle is very difficult to prove (or understand) in the case of point
particles, it is more transparent for spatially distributed entities, as in
the Dirac electron waves \citep{DuckSudarshan98}.  Specifically, for a
spatially extended spin $\frac{1}{2}$ entity with ``strings'' attached to the
surrounding space (e.g., coupling to the Maxwell field), it is clear that a
$2\pi$ rotation or an exchange of two particles (which can be envisioned as
rotating the two as a unit around their common center by $\pi$, followed by an
additional $\pi$ rotation for each individual to get them facing each other
again) leaves things still twisted, which is indicated by the asymmetric minus
sign applied for this case.  Only an additional $2\pi$ rotation (or
re-exchange) fully untwists everything.  This topological proof, originally
due to Feynman, is considered unacceptable for point particles (points don't
face in any direction for example, and don't have any obvious ``strings''),
but we would seem to avoid that problem in the distributed, coupled
Maxwell-Dirac system.  Nevertheless, actually demonstrating that the Pauli
exclusion principle emerges from this system, for example in the context of
the multi-electron interactions in He and Li atoms, would seem to be a high
priority early test for this framework.

\subsubsection{The Measurement Process}

The existing semiclassical approach provides a critical insight into the
measurement process: {\em many features of what is measured may be attributed
  to the measuring device itself, not to what is being measured!}  Thus,
instead of thinking that the EM field exhibits quantum behavior itself, we can
instead attribute the quantum behavior to the atomic system that the EM field
interacts with.  This situation is equivalent to the structure of many magic
tricks --- you are systematically misled to attribute properties to one
``obvious'' system (e.g., the rabbit that seems to disappear), when in fact
there is a less obvious system that is actually responsible (e.g., an extra
pocket in the magicians hat, that holds the rabbit hidden from view).  It
seems that perhaps people have been systematically misled by quantum magic to
attribute properties to the item being measured (e.g., a supposed ``photon''),
when in fact the measuring device is really responsible.

It seems clear that all quantum measurements involve interactions with atomic
systems, as this is what our macroscopic devices are made from.  The discrete
set of standing waves that are supported by the atomic system is the source of
the quantum nature of electromagnetic interactions.  Furthermore, atomic
systems are also involved as the {\em sources} of most things that are
measured, and it is critical to consider how this generation process shapes
the resulting waves (e.g., spontaneous emission of ``photons'' is likely to
produce discretized wave packets, instead of the plane waves often
considered in simplified analyses).

Thus, at a physical level, wave function collapse and measurement involve
waves nonlinearly interacting with quantized atomic systems, and it is this
chaotic, nonlinear interaction that creates the appearance of particle-like
properties out of the otherwise continuous wave functions.  Going back to the
polarization filters we considered earlier, the so-called wave function
collapse really amounts to a rotation of the polarization angle of the photon,
or the reflection or absorption of that photon if it doesn't make it through.
This is clearly a nonlinear interaction of the incident wave packet with the
atoms in the polarizing filter, and it can unfold directly through the
nonlinearities present in the Maxwell-Dirac system.  Similarly, the
photodetection process starts with the photoelectric effect, which also
represents a resonance dynamic between the incoming EM waves an the atomic
system, described again by (simplifications of) the Maxwell-Dirac system.
Thus, it would seem that this single coherent Maxwell-Dirac system can provide
a unified account of both state propagation and the measurement process,
thereby healing a longstanding rift in the standard QM model.

Further insight into the wave function collapse dynamics comes from analyses
undertaken within the {\em objective collapse} theories of QM, for example the
work of \citet{Pearle07,Pearle09} and \citet{GhirardiRiminiWeber86}.  They
have identified the principle of {\em gamblers ruin} as critical to deriving
the Born probability rule for a discrete measurement from the wave function.
Gamblers ruin is basically a negative feedback loop dynamic characterizing
someone who eventually loses a series of gambles, whereby the less money you
have, the lower your odds are of winning money.  Conversely, a mirror-image
positive feedback loop holds for the winner.  In the actual quantum
measurement process, this merely requires that during an iterative unfolding
dynamic measurement process the probability of increasing the wave strength
for a given measured state is proportional to its current strength.  Thus,
once one side of this tug-of-war gets a bit of an advantage, its advantage
will increase further, leading to a ``collapse'' into one alternative at the
expense of the other.  Note that the conservation property automatically
enforced by the wave function itself provides a key ``zero sum'' constraint in
this process.  This kind of dynamic can also be highly chaotic, in the sense
that small initial differences in the strength (probability) of one value in
the wave function are magnified.  Assuming a reasonable amount of variability
in the states of the wave functions for the ``identically prepared'' incoming
wave packets being measured, and of the measuring device itself, it is not
hard to see that each of the different states can be sampled effectively at
random over repeated measurements.

In summary, particle-like discrete properties should emerge from continuous
wave dynamics in the Maxwell-Dirac system through nonlinear interactions with
quantized atomic systems, providing a seemingly natural and satisfying
physical model of the measurement process.

\subsubsection{Free electrons really seem like particles}

Whereas electrons in atomic bound states really seem to behave like waves,
free electrons seem more particle-like than a pure-waves view would appear to
suggest.  They seem to be localized (e.g., Schr\"odinger's concern for the
cloud chamber tracks), they seem to have unitary charge whenever we measure
them, and they just seem to cohere as a unit more than you'd expect from
flimsy waves.  In short, if everything was just waves, you would expect lots
of splatter and extra bits of goo getting left behind all over the place,
instead of the seemingly tidy and neat, quantized behavior we seem to observe
for free electrons.

Interestingly, the first problem of dispersion of the wave packet is really a
problem for standard QM as much as it is for anything else \citep{Dorling87}.
In the specific case of the cloud chamber, it would seem that the electron is
constantly interacting with molecules in the water vapor, and this interaction
serves to constrain the packet diffusion.  In other words, there is a
continuous measurement process that imposes discretization and localization of
the electron trajectory.  In a classic quantum complementarity situation, the
wave packet only spreads when you don't measure it, and then how do you know
it has spread!?  Once you look at it, that very process of measurement
re-localizes the wave.  But somehow if the wave packet really were to get so
very widely spread out, it seems that it would be very difficult for it to not
leave some distant parts behind during the collapse process.

It would resolve a lot of problems if somehow the electron wave packet
exhibited some kind of {\em emergent localization} property, such that it
cannot spread much beyond some critical limit (e.g., the Compton wavelength).
This would have to be due to some kind of nonlinear interaction in the coupled
Maxwell-Dirac system that is not otherwise represented in the linear
Schr\"odinger wave function.  Given the spin and zitterbewegung dynamic, plus
the complex interactions this must engender with the Maxwell field, it seems
clear that the electron in this system is very much a complex dynamic entity,
which could very well exhibit this kind of emergent behavior.  Indeed,
\citet{Jaynes91} showed that the zitterbewegung property could produce a net
self-attractive force on the electron cloud, that would cause it to cohere.
Furthermore, any additional electromagnetic interactions with other electron
clouds should produce a repulsive force from the like-charges repelling, which
will help to keep the charge cloud localized.  Other analyses have shown
exponential decay in the Maxwell-Dirac density, and otherwise shown spatial
localization (e.g., solving the Cauchy problem)
\citep{Chadam73,FlatoSimonTaflin87,EstebanGeorgievSere96,Radford03}.

Another related problem is charge quantization --- why would an electron wave
always have a unitary elementary charge associated with it?  The linear
Schr\"odinger equation suffers from the unrestricted ability to perform
arbitrary scaling of the wave amplitudes, which then affects the net amount of
charge it represents.  There is nothing to fix a preferred level of charge
within a given wave packet.  But if the Jaynes and other analyses are correct,
the emergent localization property of the Maxwell-Dirac system could also fix
a given level of charge for this stable wave configuration.  Another potential
source of charge quantization is to trace it back to the strongly localized
protons in atomic nuclei.  The atomic charge then traps exactly corresponding
amounts of electronic charge clouds.  If early in the big bang everything was
tied up in Hydrogen atoms, then this would make everything initially quantized
in this way.

In summary, one could reasonably be optimistic that the various particle-like
effects can be understood as emerging from the core nonlinearities of the
Maxwell-Dirac system, plus additional constraints from the strongly localized
nuclear particles.  But clearly a huge amount of work remains to be done to
explore these possibilities.

\subsubsection{Beyond Electrodynamics}

The Dirac wave function can be configured in principle to have zero charge
value, which could potentially provide a model of the neutrino.  Furthermore,
one result from semiclassical theory is that a substantial portion of the
electron's measured mass comes from its electromagnetic self-energy
\citep[e.g.,][]{Crisp96}.  Thus, one would expect Dirac waves that do not have
charge would have considerably less mass, perhaps consistent with that of the
neutrino.  Furthermore, the other members of the lepton family (muon and tau)
could have more energetic electromagnetic dynamics, and thus a larger amount
of self-energy, producing their larger observed masses.  Thus, there is at
least the potential for an elegant way of understanding all of the members of
the lepton family (and by potential extension, the similar 3 levels of the
quarks).  Note also that there does not appear to be any need for the Higgs
mechanism to generate a particle's mass --- the increasing lack of evidence
for the Higgs boson is thus consistent with this overall framework.

Lastly, one of the major barriers to a fully unified description of all known
forces in nature, including gravitation, is that the virtual particle energy
at small length scales in the ZPF of the QED model seems incompatible with
general relativity --- spacetime would be massively warped by this field, and
if there really are random vibrations at these very high energies, then it
turns into some kind of ugly quantum spacetime foam.  To the extent that we
can eliminate the ZPF entirely within the Maxwell-Dirac framework, and explain
everything in terms of self-field effects, this would appear to be entirely
compatible with general relativity.  Indeed, it becomes just another
mutually-coupled field, and the Maxwell-Dirac-Einstein system has actually
been analyzed \citep{FinsterSmollerYau99c}, with the result that gravitation
may actually play a role in emergent localization.

\subsubsection{deBroglie-Bohm Pilot Wave Theory}

Another possible physical model of quantum mechanics is the deBroglie-Bohm
{\em pilot wave} theory, where the quantum wave function is considered real, and
it guides the behavior of the underlying particle
\citep{Bohm53,BohmHiley93,Holland93}.  Although acknowledging the realism of
the wave seems like progress, this wave remains very mysterious and seems
physical in name only --- its only purpose is to guide the behavior of the
particle (which interestingly has no effect back on the wave itself).  Because
it is formulated using linear Schr\"odinger waves (with significant
difficulties in extending to the relativistic Dirac equation), it requires
high-dimensional configuration space to describe multi-particle states, which
makes it intrinsically nonlocal.  The undefined nature of the dimensionality
of configuration space is thus a severe barrier to regarding this as a
physical model.  In short, it seems that this attempt to provide a physical
interpretation that considers both the wave and the particle to be real just
serves to highlight how nonphysical and nonlocal the standard QM calculational
model really is.

\subsubsection{Problems with Particles}

Although there are certainly some significant problems that need to be solved
for the waves-only model, it is important to give equal time to the even more
significant problems associated with any notion of particles.  Whereas the
problems with waves take the form of technical challenges and promissory
emergent properties, the problems with particles seem more fundamental.  Here
is a quick sampling:
\begin{itemize}
\item Particles are only good for creating paradoxes: The classic two-slit
  experiment is only paradoxical if you believe in particles.  There is
  absolutely nothing paradoxical about a wave packet going through both slits
  and then collapsing onto a detector somewhere to register a discrete
  detection event.  
\item How can particles ever cross nodal points in wave functions?  For
  example, we are to believe that electrons zoom around in the wave functions
  associated with the various atomic orbitals, but many of these have nodal
  points where the probability of finding an electron should be precisely 0
  --- how do they ever cross these zero points \citep{Nelson90}?
\item Point particles necessarily cause infinite self-field effects close to
  the point, requiring ugly renormalization procedures.  How would nature
  renormalize this problem away in an autonomous fashion?
\item How does one explain particle creation and destruction processes, if we
  attribute some kind of solid reality to the particle itself?  As Feynman's
  father apparently asked, ``was the photon inside the atom before it got
  emitted through spontaneous emission?''  To which Feynman had no good
  answer.  In the waves-only model (and even in QED), this is not a problem
  --- particles are just emergent wave configurations, and when these wave
  dynamics change, it appears as though particles were created or destroyed.
\item The stochastic electrodynamics (SED) model of the atomic system is
  apparently based on the premise of a point electron, which, being an
  essentially classical framework, suffers from the classical energy radiation
  problem, causing it to fall into the nucleus.  But the ZPF energy in SED is
  thought to rescue this problem, by providing a stabilizing feedback dynamic.
  However, this turns out to break down for non-circular orbits, which
  apparently fatally undermines the SED model of atomic systems
  \citep{delaPenaCetto96}.  It is not clear how such an approach would have
  dealt with the lack of angular momentum associated with electron orbitals in
  the first place, which seems to strongly undermine any kind of literal
  particle orbiting theory.  Seems that this point electron assumption may be
  more trouble than it's worth.
\end{itemize}

In summary, it seems that there are many fundamental problems associated with
particles that are obviated by the waves-only viewpoint.  If one can actually
account for all the particle-like properties from within the waves-only model,
then this seems like the best path to a paradox-free model of quantum physics.

\section{Conclusions}

By recognizing the critical differences between calculational tools and
physical models, it seems clear that quantum mechanics is completely dominated
by the tools, and suffers from the lack of physical models.  People have
systematically mistaken things like the high-dimensional configuration space
in the standard QM formalism as somehow a reflection of physical reality, when
it seems to derive instead from the kinds of simplifications (e.g., the linear
nature of the Schr\"odinger wave function) that make it an extremely useful
calculational tool.  Recent analyses suggest that the true nature of this
framework is as an abstract probability calculus, not a physical model
\citep{Hardy01,ChiribellaDArianoPerinotti11}.  Similarly, in attempting to
provide physical interpretations of the manifestly nonlocal, nonphysical
Fourier space of QED, all manner of absurdity has been promulgated
\citep{Nikolic07}, for example the idea that a frequency mode in Fourier space
corresponds to a physical entity described by the term ``photon.''  If instead
these tools were properly recognized for what they are, many layers of
confusion would be removed, and people could use these tools for all they are
worth, without suspending the quest for an underlying physical model.

Is it really possible that there is a sensible physical model for the quantum
world?  Without locality, it seems impossible.  With locality, it actually
seems relatively trivial (in the grand scheme of things) and already well
known and partially understood: the coupled Maxwell-Dirac system (for
electrodynamics).  Indeed, this system can seemingly be derived just by
systematically forbidding the usual trick of calling various things
``virtual'' that nevertheless produce actual physical effects.  The quantum
wave function must be real, because it certainly has real physical effects.
The wave equation that best describes the electron (and positron) in its full
glory is the Dirac equation.  It produces a conserved charge value.  Hence,
the electron {\em is} this charge wave.  Once the Dirac equation is coupled
with its self-field in Maxwell's equations, it becomes nonlinear, and lots of
interesting physical phenomena ensue, that have otherwise been attributed to
``virtual'' particles in QED.  Ockham's razor, plus a number of arguments as
given above, suggest that this may be as far as we need to go, at least until
proven otherwise.

Given that quantum nonlocality is the single greatest barrier to the
development of this elegant physical model, one really needs to apply an
extremely high standard of proof for this nonlocality, and deeply question
whether it is truly mandated by the physical world, or might instead be an
accident of the standard QM calculational tools.  Does the existing
experimental evidence meet this standard?  \cite{Santos05} provides a
very strong conclusion:
\begin{quote}
  In any case I claim that {\em local realism is such a fundamental principle
    that should not be dismissed without extremely strong arguments.}  It is a
  fact that there is no direct empirical evidence at all for the violation of
  local realism. The existing evidence is just that quantum mechanical
  predictions are confirmed, in general, in tests of (non-genuine Bell)
  inequalities like (18) or (14). Only when this evidence is combined with
  theoretical arguments (or prejudices) it might be argued that local realism
  is refuted. But, in my opinion, this combination is too weak for such a
  strong conclusion. Thus I propose that {\em no loophole-free experiment is
    possible which violates local realism.}
\end{quote}

But even Santos is apparently too cautious to argue that the evident success
of the semiclassical approach effectively undermines the basic predictions of
quantum mechanics regarding entangled photons in the first place.  And it is
only this entangled photon case that provides a sound basis for proving
quantum nonlocality.  If we can reject the premise that photons remain
entangled, because they are actually just classical EM waves, and we can very
reasonably question the empirical demonstrations of nonlocal photon
entanglement, then the only compelling argument left is just that QM is so
accurate in all other ways.  But it seems that entanglement is treated
inconsistently within QM (e.g., locality is required for the creation of
entanglement, but not its maintenance), and there does not seem to be a good
way to represent a formerly entangled state, which seems like a more accurate
model of photon behavior.  Putting all of this together, the case for quantum
nonlocality seems extremely shaky.  Does this shaky case really win out when
pitted against the very compelling, paradox-free physical model provided by
the Maxwell-Dirac system?  I for one would require a much stronger, iron-clad
case for nonlocality before abandoning such a promising prospect for finally
understanding the beautiful mysteries of the quantum realm.

\section{Acknowledgments}

Thanks to Emilio Santos for many useful comments on an earlier draft of this
manuscript.

\bibliographystyle{model5-names}

\begin{thebibliography}{76}
\expandafter\ifx\csname natexlab\endcsname\relax\def\natexlab#1{#1}\fi
\providecommand{\bibinfo}[2]{#2}
\ifx\xfnm\relax \def\xfnm[#1]{\unskip,\space#1}\fi
\bibitem[{Adenier \& Khrennikov(2003)}]{AdenierKhrennikov03}
\bibinfo{author}{Adenier, G.}, \& \bibinfo{author}{Khrennikov, A.~Y.}
  (\bibinfo{year}{2003}).
\newblock \bibinfo{title}{Testing the fair sampling assumption for epr-bell
  experiments with polarizer beamsplitters}.
\newblock \bibinfo{note}{ArXiv:quant-ph/0306045v6}.
\bibitem[{Adenier \& Khrennikov(2007)}]{AdenierKhrennikov07}
\bibinfo{author}{Adenier, G.}, \& \bibinfo{author}{Khrennikov, A.~Y.}
  (\bibinfo{year}{2007}).
\newblock \bibinfo{title}{Is the fair sampling assumption supported by epr
  experiments?}
\newblock {\it \bibinfo{journal}{Journal of Physics B: Atomic, Molecular and
  Optical Physics}\/},  {\it \bibinfo{volume}{40}\/}, \bibinfo{pages}{131}.
\bibitem[{Akbar~Ali et~al.(2010)Akbar~Ali, Bishit, Nautiyal, Shukla, Bindra \&
  Oak}]{AkbarAliEtAl10}
\bibinfo{author}{Akbar~Ali, S.}, \bibinfo{author}{Bishit, P.~B.},
  \bibinfo{author}{Nautiyal, A.}, \bibinfo{author}{Shukla, V.},
  \bibinfo{author}{Bindra, K.~S.}, \& \bibinfo{author}{Oak, S.~M.}
  (\bibinfo{year}{2010}).
\newblock \bibinfo{title}{Conical emission in beta-barium borate under
  femtosecond pumping with phase matching angles away from second harmonic
  generation}.
\newblock {\it \bibinfo{journal}{Journal of the Optical Society of America
  B.}\/},  {\it \bibinfo{volume}{27}\/}, \bibinfo{pages}{1751--1756}.
\bibitem[{Argaman \& Makov(2000)}]{ArgamanMakov00}
\bibinfo{author}{Argaman, N.}, \& \bibinfo{author}{Makov, G.}
  (\bibinfo{year}{2000}).
\newblock \bibinfo{title}{Density functional theory: An introduction}.
\newblock {\it \bibinfo{journal}{American Journal of Physics}\/},  {\it
  \bibinfo{volume}{68}\/}, \bibinfo{pages}{69--79}.
\bibitem[{Aschwanden et~al.(2006)Aschwanden, Philipp, Hess, Barraza-Lopez \&
  Adenier}]{AschwandenEtAl06}
\bibinfo{author}{Aschwanden, M.}, \bibinfo{author}{Philipp, W.},
  \bibinfo{author}{Hess, K.}, \bibinfo{author}{Barraza-Lopez, S.}, \&
  \bibinfo{author}{Adenier, G.} (\bibinfo{year}{2006}).
\newblock \bibinfo{title}{Local time dependent instruction-set model for the
  experiment of pan et al}.
\newblock {\it \bibinfo{journal}{Quantum Theory: Reconsideration of
  Foundations}\/},  {\it \bibinfo{volume}{810}\/}, \bibinfo{pages}{437--446}.
\bibitem[{Aspect et~al.(1982{\natexlab{a}})Aspect, Dalibard \&
  Roger}]{AspectDalibardRoger82}
\bibinfo{author}{Aspect, A.}, \bibinfo{author}{Dalibard, J.}, \&
  \bibinfo{author}{Roger, G.} (\bibinfo{year}{1982}{\natexlab{a}}).
\newblock \bibinfo{title}{Experimental test of bell's inequalities using
  time-varying analyzers}.
\newblock {\it \bibinfo{journal}{Physical Review Letters}\/},  {\it
  \bibinfo{volume}{49}\/}, \bibinfo{pages}{1804--1807}.
\bibitem[{Aspect et~al.(1982{\natexlab{b}})Aspect, Grainger \&
  Roger}]{AspectGraingerRoger82}
\bibinfo{author}{Aspect, A.}, \bibinfo{author}{Grainger, P.}, \&
  \bibinfo{author}{Roger, G.} (\bibinfo{year}{1982}{\natexlab{b}}).
\newblock \bibinfo{title}{Experimental realization of
  einstein-podolsky-rosen-bohm gedankenexperiment: A new violation of bell's
  inequalities}.
\newblock {\it \bibinfo{journal}{Physical Review Letters}\/},  {\it
  \bibinfo{volume}{49}\/}, \bibinfo{pages}{91--94}.
\bibitem[{Barut(1991)}]{Barut91}
\bibinfo{author}{Barut, A.~O.} (\bibinfo{year}{1991}).
\newblock \bibinfo{title}{Brief history and recent developments in electron
  theory and quantumelectrodynamics}.
\newblock In \bibinfo{editor}{D.~Hestenes}, \&
  \bibinfo{editor}{A.~Weingartshofer} (Eds.), {\it \bibinfo{booktitle}{The
  Electron: New Theory and Experiment}\/} (pp. \bibinfo{pages}{105--148}).
\newblock \bibinfo{address}{Dordrecht}: \bibinfo{publisher}{Kluwer Academic}.
\bibitem[{Bell(1964)}]{Bell64}
\bibinfo{author}{Bell, J.~S.} (\bibinfo{year}{1964}).
\newblock \bibinfo{title}{On the {Einstein-Podolsky-Rosen} paradox}.
\newblock {\it \bibinfo{journal}{Physics}\/},  {\it \bibinfo{volume}{1}\/},
  \bibinfo{pages}{195--200}.
\bibitem[{Bell(1987)}]{Bell87}
\bibinfo{author}{Bell, J.~S.} (\bibinfo{year}{1987}).
\newblock {\it \bibinfo{title}{Speakable and unspeakable in quantum mechanics
  (collected papers in quantum philosophy)}\/}.
\newblock \bibinfo{address}{Cambridge}: \bibinfo{publisher}{Cambridge
  University Press}.
\bibitem[{Bialynicki-Birula(1994)}]{BialynickiBirula94}
\bibinfo{author}{Bialynicki-Birula, I.} (\bibinfo{year}{1994}).
\newblock \bibinfo{title}{{Weyl}, {Dirac}, and {Maxwell} equations on a lattice
  as unitary cellular automata}.
\newblock {\it \bibinfo{journal}{Physical Review D}\/},  {\it
  \bibinfo{volume}{49}\/}, \bibinfo{pages}{6920--6927}.
\bibitem[{Bohm(1953)}]{Bohm53}
\bibinfo{author}{Bohm, D.} (\bibinfo{year}{1953}).
\newblock \bibinfo{title}{Proof that probability density approaches j/j 2 in
  causal interpretation of quantum theory.}
\newblock {\it \bibinfo{journal}{Physical Review}\/},  {\it
  \bibinfo{volume}{89}\/}, \bibinfo{pages}{458--466}.
\bibitem[{Bohm \& Hiley(1993)}]{BohmHiley93}
\bibinfo{author}{Bohm, D.}, \& \bibinfo{author}{Hiley, B.~J.}
  (\bibinfo{year}{1993}).
\newblock {\it \bibinfo{title}{The Unidivided Universe: An Ontological
  Interpretation of Quantum Theory}\/}.
\newblock \bibinfo{address}{London}: \bibinfo{publisher}{Routledge}.
\bibitem[{Brill \& Goodman(1967)}]{BrillGoodman67}
\bibinfo{author}{Brill, O.~L.}, \& \bibinfo{author}{Goodman, B.}
  (\bibinfo{year}{1967}).
\newblock \bibinfo{title}{Causality in the coulomb gauge}.
\newblock {\it \bibinfo{journal}{American Journal of Physics}\/},  {\it
  \bibinfo{volume}{35}\/}, \bibinfo{pages}{832--837}.
\bibitem[{Chadam(1973)}]{Chadam73}
\bibinfo{author}{Chadam, J.~M.} (\bibinfo{year}{1973}).
\newblock \bibinfo{title}{Global solutions of the cauchy problem for the
  (classical) coupled maxwell-dirac equations in one space dimension}.
\newblock {\it \bibinfo{journal}{Journal of Functional Analysis}\/},  {\it
  \bibinfo{volume}{13}\/}, \bibinfo{pages}{173184}.
\bibitem[{Chiribella et~al.(2011)Chiribella, D'Ariano \&
  Perinotti}]{ChiribellaDArianoPerinotti11}
\bibinfo{author}{Chiribella, G.}, \bibinfo{author}{D'Ariano, G.~M.}, \&
  \bibinfo{author}{Perinotti, P.} (\bibinfo{year}{2011}).
\newblock \bibinfo{title}{Informational derivation of quantum theory}.
\newblock {\it \bibinfo{journal}{Physical Review A}\/},  {\it
  \bibinfo{volume}{84}\/}, \bibinfo{pages}{012311}.
\bibitem[{Crisp(1996)}]{Crisp96}
\bibinfo{author}{Crisp, M.~D.} (\bibinfo{year}{1996}).
\newblock \bibinfo{title}{Relativistic neoclassical radiation theory}.
\newblock {\it \bibinfo{journal}{Physical Review A}\/},  {\it
  \bibinfo{volume}{54}\/}, \bibinfo{pages}{87--92}.
\bibitem[{Dorling(1987)}]{Dorling87}
\bibinfo{author}{Dorling, J.} (\bibinfo{year}{1987}).
\newblock \bibinfo{title}{Schrodinger's original interpretation of the
  schrodinger equation: A rescue attempt}.
\newblock In \bibinfo{editor}{C.~W. Klimister} (Ed.), {\it
  \bibinfo{booktitle}{Schrodinger: Centenary Celebration of a Polymath}\/} (pp.
  \bibinfo{pages}{16--40}).
\newblock \bibinfo{address}{Cambridge}: \bibinfo{publisher}{Cambridge
  University Press}.
\bibitem[{Duck \& Sudarshan(1998)}]{DuckSudarshan98}
\bibinfo{author}{Duck, I.}, \& \bibinfo{author}{Sudarshan, E. C.~G.}
  (\bibinfo{year}{1998}).
\newblock \bibinfo{title}{Toward and understanding of the spin-statistics
  theorem}.
\newblock {\it \bibinfo{journal}{American Journal of Physics}\/},  {\it
  \bibinfo{volume}{66}\/}, \bibinfo{pages}{284--303}.
\bibitem[{Einstein et~al.(1935)Einstein, Podolsky \&
  Rosen}]{EinsteinPodolskyRosen35}
\bibinfo{author}{Einstein, A.}, \bibinfo{author}{Podolsky, B.}, \&
  \bibinfo{author}{Rosen, N.} (\bibinfo{year}{1935}).
\newblock \bibinfo{title}{Can quantum-mechanical description of physical
  reality be considered complete?}
\newblock {\it \bibinfo{journal}{Physical Review}\/},  {\it
  \bibinfo{volume}{47}\/}, \bibinfo{pages}{777--780}.
\bibitem[{Esteban et~al.(1996)Esteban, Georgiev \&
  Sere}]{EstebanGeorgievSere96}
\bibinfo{author}{Esteban, M.~J.}, \bibinfo{author}{Georgiev, V.}, \&
  \bibinfo{author}{Sere, E.} (\bibinfo{year}{1996}).
\newblock \bibinfo{title}{Stationary solutions of the maxwell-dirac and the
  klein-gordon-dirac equations}.
\newblock {\it \bibinfo{journal}{Calculus of Variations and Partial
  Differential Equations}\/},  {\it \bibinfo{volume}{4}\/},
  \bibinfo{pages}{265--281}.
\bibitem[{Finster et~al.(1999)Finster, Smoller \& Yau}]{FinsterSmollerYau99c}
\bibinfo{author}{Finster, F.}, \bibinfo{author}{Smoller, J.}, \&
  \bibinfo{author}{Yau, S.-T.} (\bibinfo{year}{1999}).
\newblock \bibinfo{title}{The coupling of gravity to spin and
  electromagnetism}.
\newblock {\it \bibinfo{journal}{Modern Physics Letters A}\/},  {\it
  \bibinfo{volume}{14}\/}, \bibinfo{pages}{1053--1057}.
\bibitem[{Flato et~al.(1987)Flato, Simon \& Taflin}]{FlatoSimonTaflin87}
\bibinfo{author}{Flato, M.}, \bibinfo{author}{Simon, J.}, \&
  \bibinfo{author}{Taflin, E.} (\bibinfo{year}{1987}).
\newblock \bibinfo{title}{On the global solutions of the maxwell-dirac
  equations}.
\newblock {\it \bibinfo{journal}{Comm. Math. Physics}\/},  {\it
  \bibinfo{volume}{113}\/}, \bibinfo{pages}{21--49}.
\bibitem[{Fredkin(1990)}]{Fredkin90}
\bibinfo{author}{Fredkin, E.} (\bibinfo{year}{1990}).
\newblock \bibinfo{title}{Digital mechanics}.
\newblock {\it \bibinfo{journal}{Physica D}\/},  {\it \bibinfo{volume}{45}\/},
  \bibinfo{pages}{254--270}.
\bibitem[{Fredkin \& Toffoli(1982)}]{FredkinToffoli82}
\bibinfo{author}{Fredkin, E.}, \& \bibinfo{author}{Toffoli, T.}
  (\bibinfo{year}{1982}).
\newblock \bibinfo{title}{Conservative logic}.
\newblock {\it \bibinfo{journal}{International Journal of Theoretical
  Physics}\/},  {\it \bibinfo{volume}{21}\/}, \bibinfo{pages}{219--253}.
\bibitem[{Gardner(1970)}]{Gardner70}
\bibinfo{author}{Gardner, M.} (\bibinfo{year}{1970}).
\newblock \bibinfo{title}{Mathematical games: The fantastic combinations of
  {\em john conway's} new solitare game ``life''}.
\newblock {\it \bibinfo{journal}{Scientific American}\/},  {\it
  \bibinfo{volume}{223}\/}, \bibinfo{pages}{120--123}.
\bibitem[{Gerry \& Knight(2005)}]{GerryKnight05}
\bibinfo{author}{Gerry, C.~C.}, \& \bibinfo{author}{Knight, P.~L.}
  (\bibinfo{year}{2005}).
\newblock {\it \bibinfo{title}{Introductory Quantum Optics}\/}.
\newblock \bibinfo{address}{Cambridge}: \bibinfo{publisher}{Cambridge
  University Press}.
\bibitem[{Ghirardi et~al.(1986)Ghirardi, Rimini \&
  Weber}]{GhirardiRiminiWeber86}
\bibinfo{author}{Ghirardi, G.~C.}, \bibinfo{author}{Rimini, A.}, \&
  \bibinfo{author}{Weber, T.} (\bibinfo{year}{1986}).
\newblock \bibinfo{title}{Unified dynamics for microscopic and macroscopic
  systems}.
\newblock {\it \bibinfo{journal}{Physical Review D}\/},  {\it
  \bibinfo{volume}{34}\/}, \bibinfo{pages}{470--491}.
\bibitem[{Grainger et~al.(1986)Grainger, Roger \&
  Aspect}]{GraingerRogerAspect86}
\bibinfo{author}{Grainger, P.}, \bibinfo{author}{Roger, G.}, \&
  \bibinfo{author}{Aspect, A.} (\bibinfo{year}{1986}).
\newblock \bibinfo{title}{Experimental evidence for a photon anticorrelation
  effect on a beam splitter: A new light on single-photon interferences}.
\newblock {\it \bibinfo{journal}{Europhysics Letters}\/},  {\it
  \bibinfo{volume}{1}\/}, \bibinfo{pages}{173--179}.
\bibitem[{Grandy(1991)}]{Grandy91}
\bibinfo{author}{Grandy, W.~T.} (\bibinfo{year}{1991}).
\newblock \bibinfo{title}{The explicit nonlinearity of quantum
  electrodynamics}.
\newblock In \bibinfo{editor}{D.~Hestenes}, \&
  \bibinfo{editor}{A.~Weingartshofer} (Eds.), {\it \bibinfo{booktitle}{The
  Electron: New Theory and Experiment}\/} (pp. \bibinfo{pages}{149--164}).
\newblock \bibinfo{address}{Dordrecht}: \bibinfo{publisher}{Kluwer Academic}.
\bibitem[{Gudder(1970)}]{Gudder70}
\bibinfo{author}{Gudder, S.~P.} (\bibinfo{year}{1970}).
\newblock \bibinfo{title}{On hidden-variable theories}.
\newblock {\it \bibinfo{journal}{Journal of Mathematical Physics}\/},  {\it
  \bibinfo{volume}{11}\/}, \bibinfo{pages}{431--436}.
\bibitem[{Hardy(2001)}]{Hardy01}
\bibinfo{author}{Hardy, L.} (\bibinfo{year}{2001}).
\newblock \bibinfo{title}{Quantum theory from five reasonable axioms}.
\newblock \bibinfo{note}{ArXiv:quant-ph/0101012v4}.
\bibitem[{Holland(1993)}]{Holland93}
\bibinfo{author}{Holland, P.~R.} (\bibinfo{year}{1993}).
\newblock {\it \bibinfo{title}{The Quantum Theory of Motion: An Account of the
  de Broglie-Bohm Causal Interpretation of Quantum Mechanics}\/}.
\newblock \bibinfo{address}{Cambridge}: \bibinfo{publisher}{Cambridge
  University Press}.
\bibitem[{Hong et~al.(1987)Hong, Ou \& Mandel}]{HongOuMandel87}
\bibinfo{author}{Hong, C.~K.}, \bibinfo{author}{Ou, Z.~Y.}, \&
  \bibinfo{author}{Mandel, L.} (\bibinfo{year}{1987}).
\newblock \bibinfo{title}{Measurement of subpicosecond time intervals between
  two photons by interference}.
\newblock {\it \bibinfo{journal}{Physical Review Letters}\/},  {\it
  \bibinfo{volume}{59}\/}, \bibinfo{pages}{2044--2046}.
\bibitem[{Jackson(2002)}]{Jackson02}
\bibinfo{author}{Jackson, J.~D.} (\bibinfo{year}{2002}).
\newblock \bibinfo{title}{From lorenz to coulomb and other explicit gauge
  transformations}.
\newblock {\it \bibinfo{journal}{American Journal of Physics}\/},  {\it
  \bibinfo{volume}{70}\/}, \bibinfo{pages}{917--928}.
\bibitem[{Jaynes(1973)}]{Jaynes73}
\bibinfo{author}{Jaynes, E.~T.} (\bibinfo{year}{1973}).
\newblock \bibinfo{title}{Survey of the present status of neoclassical
  radiation theory}.
\newblock In \bibinfo{editor}{L.~Mandel}, \& \bibinfo{editor}{E.~Wolf} (Eds.),
  {\it \bibinfo{booktitle}{Coherence and Quantum Optics}\/} (pp.
  \bibinfo{pages}{35--81}).
\newblock \bibinfo{address}{New York}: \bibinfo{publisher}{Plenum Press}.
\bibitem[{Jaynes(1978)}]{Jaynes78}
\bibinfo{author}{Jaynes, E.~T.} (\bibinfo{year}{1978}).
\newblock \bibinfo{title}{Electrodynamics today}.
\newblock In \bibinfo{editor}{L.~Mandel}, \& \bibinfo{editor}{E.~Wolf} (Eds.),
  {\it \bibinfo{booktitle}{Coherence and Quantum Optics {IV}}\/} (pp.
  \bibinfo{pages}{495--509}).
\newblock \bibinfo{address}{New York}: \bibinfo{publisher}{Plenum Press}.
\bibitem[{Jaynes(1991)}]{Jaynes91}
\bibinfo{author}{Jaynes, E.~T.} (\bibinfo{year}{1991}).
\newblock \bibinfo{title}{Scattering of light by free eelctrons as a test of
  quantum theory}.
\newblock In \bibinfo{editor}{D.~Hestenes}, \&
  \bibinfo{editor}{A.~Weingartshofer} (Eds.), {\it \bibinfo{booktitle}{The
  Electron: New Theory and Experiment}\/} (pp. \bibinfo{pages}{1--20}).
\newblock \bibinfo{address}{Dordrecht}: \bibinfo{publisher}{Kluwer Academic}.
\bibitem[{Jaynes \& Cummings(1963)}]{JaynesCummings63}
\bibinfo{author}{Jaynes, E.~T.}, \& \bibinfo{author}{Cummings, F.~W.}
  (\bibinfo{year}{1963}).
\newblock {\it \bibinfo{journal}{Proceedings of the IEEE}\/},  {\it
  \bibinfo{volume}{51}\/}, \bibinfo{pages}{89}.
\bibitem[{Khrennikov(2001)}]{Khrennikov01}
\bibinfo{author}{Khrennikov, A.} (\bibinfo{year}{2001}).
\newblock \bibinfo{title}{Linear representations of probabilistic
  transformations induced by context transitions}.
\newblock {\it \bibinfo{journal}{Journal of Physics A: Mathematical and
  General}\/},  {\it \bibinfo{volume}{34}\/}, \bibinfo{pages}{9965--9981}.
\bibitem[{Klassen(2011)}]{Klassen11}
\bibinfo{author}{Klassen, S.} (\bibinfo{year}{2011}).
\newblock \bibinfo{title}{The photoelectric effect: Reconstructing the story
  for the physics classroom}.
\newblock {\it \bibinfo{journal}{Science and Education}\/},  {\it
  \bibinfo{volume}{20}\/}, \bibinfo{pages}{719--731}.
\bibitem[{Lamb(1995)}]{Lamb95}
\bibinfo{author}{Lamb, W.~E.} (\bibinfo{year}{1995}).
\newblock \bibinfo{title}{Anti-photon}.
\newblock {\it \bibinfo{journal}{Applied Physics B: Lasers and Optics}\/},
  {\it \bibinfo{volume}{60}\/}, \bibinfo{pages}{77--84}.
\bibitem[{{Mandel}(1976)}]{Mandel76}
\bibinfo{author}{{Mandel}, L.} (\bibinfo{year}{1976}).
\newblock \bibinfo{title}{The case for and against semiclassical radiation
  theory}.
\newblock {\it \bibinfo{journal}{Progress in optics}\/},  {\it
  \bibinfo{volume}{13}\/}, \bibinfo{pages}{27--68}.
\bibitem[{Marshall \& Santos(1988)}]{MarshallSantos88}
\bibinfo{author}{Marshall, T.}, \& \bibinfo{author}{Santos, E.}
  (\bibinfo{year}{1988}).
\newblock \bibinfo{title}{Stochastic optics: A reaffirmation of the wave nature
  of light}.
\newblock {\it \bibinfo{journal}{Foundations of Physics}\/},  {\it
  \bibinfo{volume}{18}\/}, \bibinfo{pages}{185--223}.
\bibitem[{Marshall(2002)}]{Marshall02}
\bibinfo{author}{Marshall, T.~W.} (\bibinfo{year}{2002}).
\newblock \bibinfo{title}{Nonlocality -- the party may be over!}
\newblock \bibinfo{note}{ArXiv:quant-ph/0203042v1}.
\bibitem[{Marshall \& Santos(1985)}]{MarshallSantos85}
\bibinfo{author}{Marshall, T.~W.}, \& \bibinfo{author}{Santos, E.}
  (\bibinfo{year}{1985}).
\newblock \bibinfo{title}{Local realist model for the coincidence rates in
  atomic cascade experiments}.
\newblock {\it \bibinfo{journal}{Physics Letters}\/},  {\it
  \bibinfo{volume}{107}\/}, \bibinfo{pages}{164--168}.
\bibitem[{Marshall \& Santos(1997)}]{MarshallSantos97}
\bibinfo{author}{Marshall, T.~W.}, \& \bibinfo{author}{Santos, E.}
  (\bibinfo{year}{1997}).
\newblock \bibinfo{title}{The myth of the photon}.
\newblock In \bibinfo{editor}{S.~Jeffers} (Ed.), {\it \bibinfo{booktitle}{The
  Present Status of the Quantum Theory of Light}\/} (pp.
  \bibinfo{pages}{67--77}).
\newblock \bibinfo{address}{Dordrecht}: \bibinfo{publisher}{Kluwer Academic}.
\bibitem[{Marshall et~al.(1983)Marshall, Santos \&
  Selleri}]{MarshallSantosSelleri83}
\bibinfo{author}{Marshall, T.~W.}, \bibinfo{author}{Santos, E.}, \&
  \bibinfo{author}{Selleri, F.} (\bibinfo{year}{1983}).
\newblock \bibinfo{title}{Local realism has not been refuted by atomic cascade
  experiments}.
\newblock {\it \bibinfo{journal}{Physics Letters}\/},  {\it
  \bibinfo{volume}{98}\/}, \bibinfo{pages}{5--9}.
\bibitem[{Meyer(1996)}]{Meyer96}
\bibinfo{author}{Meyer, D.~A.} (\bibinfo{year}{1996}).
\newblock \bibinfo{title}{From quantum cellular automata to quantum lattice
  gasses}.
\newblock {\it \bibinfo{journal}{Journal of Statistical Physics}\/},  {\it
  \bibinfo{volume}{85}\/}, \bibinfo{pages}{551--574}.
\bibitem[{Milonni(1984)}]{Milonni84}
\bibinfo{author}{Milonni, P.~W.} (\bibinfo{year}{1984}).
\newblock \bibinfo{title}{Why spontaneous emission?}
\newblock {\it \bibinfo{journal}{American Journal of Physics}\/},  {\it
  \bibinfo{volume}{52}\/}, \bibinfo{pages}{340--343}.
\bibitem[{Nelson(1990)}]{Nelson90}
\bibinfo{author}{Nelson, P.~G.} (\bibinfo{year}{1990}).
\newblock \bibinfo{title}{How do electrons get across nodes?}
\newblock {\it \bibinfo{journal}{Journal of Chemical Education}\/},  {\it
  \bibinfo{volume}{67}\/}, \bibinfo{pages}{643--647}.
\bibitem[{Nikolic(2007)}]{Nikolic07}
\bibinfo{author}{Nikolic, H.} (\bibinfo{year}{2007}).
\newblock \bibinfo{title}{Quantum mechanics: Myths and facts}.
\newblock {\it \bibinfo{journal}{Foundations of Physics}\/},  {\it
  \bibinfo{volume}{37}\/}, \bibinfo{pages}{1563--1611}.
\bibitem[{Onoochin(2002)}]{Onoochin02}
\bibinfo{author}{Onoochin, V.~V.} (\bibinfo{year}{2002}).
\newblock \bibinfo{title}{On non-equivalence of lorenz and coulomb gauges
  within classical electrodynamics}.
\newblock {\it \bibinfo{journal}{Annales de la Fondation Louis de Broglie}\/},
  {\it \bibinfo{volume}{27}\/}, \bibinfo{pages}{163--184}.
\bibitem[{Pearle(2007)}]{Pearle07}
\bibinfo{author}{Pearle, P.} (\bibinfo{year}{2007}).
\newblock \bibinfo{title}{How stands collapse i}.
\newblock {\it \bibinfo{journal}{Journal of Physics A: Mathematical and
  Theoretical}\/},  {\it \bibinfo{volume}{40}\/}, \bibinfo{pages}{3189--3204}.
\bibitem[{Pearle(2009)}]{Pearle09}
\bibinfo{author}{Pearle, P.} (\bibinfo{year}{2009}).
\newblock \bibinfo{title}{How stands collapse ii}.
\newblock In {\it \bibinfo{booktitle}{Quantum Reality, Relativistic Causality,
  and Closing the Epistemic Circle}\/} (pp. \bibinfo{pages}{257--292}).
\newblock \bibinfo{publisher}{Springer}.
\bibitem[{de~la Pena \& Cetto(1996)}]{delaPenaCetto96}
\bibinfo{author}{de~la Pena, L.}, \& \bibinfo{author}{Cetto, A.~M.}
  (\bibinfo{year}{1996}).
\newblock {\it \bibinfo{title}{The Quantum Dice: An Introduction to Stochastic
  Electrodynamics}\/}.
\newblock \bibinfo{address}{Dordrecht}: \bibinfo{publisher}{Kluwer Academic
  Publishers}.
\bibitem[{Poundstone(1985)}]{Poundstone85}
\bibinfo{author}{Poundstone, W.} (\bibinfo{year}{1985}).
\newblock {\it \bibinfo{title}{The Recrusive Universe}\/}.
\newblock \bibinfo{address}{Chicago, IL}: \bibinfo{publisher}{Contemporary
  Books}.
\bibitem[{Radford(2003)}]{Radford03}
\bibinfo{author}{Radford, C.~J.} (\bibinfo{year}{2003}).
\newblock \bibinfo{title}{The stationary {Maxwell-Dirac} equations}.
\newblock {\it \bibinfo{journal}{Journal of Physics A: Mathematical and
  General}\/},  {\it \bibinfo{volume}{36}\/}, \bibinfo{pages}{5663--5681}.
\bibitem[{Rovelli(1996)}]{Rovelli96}
\bibinfo{author}{Rovelli, C.} (\bibinfo{year}{1996}).
\newblock \bibinfo{title}{Relational quantum mechanics}.
\newblock {\it \bibinfo{journal}{International Journal of Theoretical
  Physics}\/},  {\it \bibinfo{volume}{35}\/}, \bibinfo{pages}{1637--1678}.
\bibitem[{Rowe et~al.(2001)Rowe, Kielpinski, Meyer, Sackett, Itano, Monroe \&
  Wineland}]{RoweEtAl01}
\bibinfo{author}{Rowe, M.~A.}, \bibinfo{author}{Kielpinski, D.},
  \bibinfo{author}{Meyer, V.}, \bibinfo{author}{Sackett, C.~A.},
  \bibinfo{author}{Itano, W.~M.}, \bibinfo{author}{Monroe, C.}, \&
  \bibinfo{author}{Wineland, D.~J.} (\bibinfo{year}{2001}).
\newblock \bibinfo{title}{Experimental violation of a bell's inequality with
  efficient detection}.
\newblock {\it \bibinfo{journal}{Nature}\/},  {\it \bibinfo{volume}{409}\/},
  \bibinfo{pages}{791--794}.
\bibitem[{Roychoudhuri \& Roy(2003)}]{RoychoudhuriRoy03}
\bibinfo{author}{Roychoudhuri, C.}, \& \bibinfo{author}{Roy, R.}
  (\bibinfo{year}{2003}).
\newblock \bibinfo{title}{The nature of light: What is a photon?}
\newblock {\it \bibinfo{journal}{Optics and Photonics News: Trends}\/},  {\it
  \bibinfo{volume}{14}\/}, \bibinfo{pages}{1--36}.
\bibitem[{Santos(2004)}]{Santos04}
\bibinfo{author}{Santos, E.} (\bibinfo{year}{2004}).
\newblock \bibinfo{title}{The failure to perform a loophole-free test of bell's
  inequality supports local realism}.
\newblock {\it \bibinfo{journal}{Foundations of Physics}\/},  {\it
  \bibinfo{volume}{34}\/}, \bibinfo{pages}{1643--1673}.
\bibitem[{Santos(2005)}]{Santos05}
\bibinfo{author}{Santos, E.} (\bibinfo{year}{2005}).
\newblock \bibinfo{title}{Bell's theorem and the experiments: Increasing
  empirical support for local realism?}
\newblock {\it \bibinfo{journal}{Studies in History and Philosophy of Modern
  Physics}\/},  {\it \bibinfo{volume}{36}\/}, \bibinfo{pages}{544--565}.
\bibitem[{Santos(2009)}]{Santos09}
\bibinfo{author}{Santos, E.} (\bibinfo{year}{2009}).
\newblock \bibinfo{title}{Relevance of a random choice in tests of bell
  inequalities with two atomic qubits}.
\newblock {\it \bibinfo{journal}{Physical Review A}\/},  {\it
  \bibinfo{volume}{79}\/}, \bibinfo{pages}{044104}.
\bibitem[{Shimony(1984)}]{Shimony84}
\bibinfo{author}{Shimony, A.} (\bibinfo{year}{1984}).
\newblock \bibinfo{title}{Contextual hidden variables theories and bell's
  inequalities}.
\newblock {\it \bibinfo{journal}{British Journal for the Philosophy of
  Science}\/},  {\it \bibinfo{volume}{35}\/}, \bibinfo{pages}{25--45}.
\bibitem[{Sipe(1995)}]{Sipe95}
\bibinfo{author}{Sipe, J.~E.} (\bibinfo{year}{1995}).
\newblock \bibinfo{title}{Photon wave functions}.
\newblock {\it \bibinfo{journal}{Physical Review A}\/},  {\it
  \bibinfo{volume}{52}\/}, \bibinfo{pages}{1875--1883}.
\bibitem[{Sulcs(2003)}]{Sulcs03}
\bibinfo{author}{Sulcs, S.} (\bibinfo{year}{2003}).
\newblock \bibinfo{title}{The nature of light and twentieth century
  experimental physics}.
\newblock {\it \bibinfo{journal}{Foundations of Science}\/},  {\it
  \bibinfo{volume}{8}\/}, \bibinfo{pages}{365--391}.
\bibitem[{Sulcs \& Osborne(2002)}]{SulcsOsborne02}
\bibinfo{author}{Sulcs, S.}, \& \bibinfo{author}{Osborne, C.~F.}
  (\bibinfo{year}{2002}).
\newblock \bibinfo{title}{Computer simulation of photon anticorrelation
  expriment using additive pre-detection noise and finite instrument
  bandwidth}.
\newblock {\it \bibinfo{journal}{International Journal of Modern Physics C
  (IJMPC)}\/},  {\it \bibinfo{volume}{13}\/}, \bibinfo{pages}{823--828}.
\bibitem[{Sun et~al.(2009)Sun, Zhang, Jia, Wang \& Sun}]{SunZhangJiaEtAl09}
\bibinfo{author}{Sun, J.}, \bibinfo{author}{Zhang, S.}, \bibinfo{author}{Jia,
  T.}, \bibinfo{author}{Wang, Z.}, \& \bibinfo{author}{Sun, Z.}
  (\bibinfo{year}{2009}).
\newblock \bibinfo{title}{Femtosecond spontaneous parametric upconversion and
  downconversion in a quadratic nonlinear medium}.
\newblock {\it \bibinfo{journal}{Journal of the Optical Society of America
  B.}\/},  {\it \bibinfo{volume}{26}\/}, \bibinfo{pages}{549--553}.
\bibitem[{Thompson(1996)}]{Thompson96}
\bibinfo{author}{Thompson, C.} (\bibinfo{year}{1996}).
\newblock \bibinfo{title}{The chaotic ball: An intuitive analogy for epr
  experiments}.
\newblock {\it \bibinfo{journal}{Foundations of Physics Letters}\/},  {\it
  \bibinfo{volume}{9}\/}, \bibinfo{pages}{357}.
\bibitem[{Tittel et~al.(1998)Tittel, Brendel, Gisin, Herzog, Zbinden \&
  Gisin}]{TittelEtAl98}
\bibinfo{author}{Tittel, W.}, \bibinfo{author}{Brendel, J.},
  \bibinfo{author}{Gisin, B.}, \bibinfo{author}{Herzog, T.},
  \bibinfo{author}{Zbinden, H.}, \& \bibinfo{author}{Gisin, N.}
  (\bibinfo{year}{1998}).
\newblock \bibinfo{title}{Experimental demonstration of quantum-correlations
  over more than 10 kilometers}.
\newblock {\it \bibinfo{journal}{Physical Review A}\/},  {\it
  \bibinfo{volume}{57}\/}, \bibinfo{pages}{3229}.
\bibitem[{Toffoli \& Margolus(1990)}]{ToffoliMargolus90}
\bibinfo{author}{Toffoli, T.}, \& \bibinfo{author}{Margolus, N.}
  (\bibinfo{year}{1990}).
\newblock \bibinfo{title}{Invertible cellular automata: A review}.
\newblock {\it \bibinfo{journal}{Physica D}\/},  {\it \bibinfo{volume}{45}\/},
  \bibinfo{pages}{229--253}.
\bibitem[{Ulam(1952)}]{Ulam50}
\bibinfo{author}{Ulam, S.} (\bibinfo{year}{1952}).
\newblock \bibinfo{title}{Random processes and transformations}.
\newblock {\it \bibinfo{journal}{Proceedings of the International Congress on
  Mathematics, 1950}\/},  {\it \bibinfo{volume}{2}\/},
  \bibinfo{pages}{264--275}.
\bibitem[{Wolfram(1983)}]{Wolfram83}
\bibinfo{author}{Wolfram, S.} (\bibinfo{year}{1983}).
\newblock \bibinfo{title}{Statistical mechanics of cellular automata}.
\newblock {\it \bibinfo{journal}{Review of Modern Physics}\/},  {\it
  \bibinfo{volume}{55}\/}, \bibinfo{pages}{601--644}.
\bibitem[{Zhang et~al.(2011)Zhang, Chen, Liu, Loy, Wong \& Du}]{ZhangEtAl11}
\bibinfo{author}{Zhang, S.}, \bibinfo{author}{Chen, J.~F.},
  \bibinfo{author}{Liu, C.}, \bibinfo{author}{Loy, M. M.~T.},
  \bibinfo{author}{Wong, G. K.~L.}, \& \bibinfo{author}{Du, S.}
  (\bibinfo{year}{2011}).
\newblock \bibinfo{title}{Optical precursor of a single photon}.
\newblock {\it \bibinfo{journal}{Phys. Rev. Lett.}\/},  {\it
  \bibinfo{volume}{106}\/}, \bibinfo{pages}{243602}.
\bibitem[{Zuse(1970)}]{Zuse70}
\bibinfo{author}{Zuse, K.} (\bibinfo{year}{1970}).
\newblock {\it \bibinfo{title}{Calculating Space}\/}.
\newblock \bibinfo{address}{Cambridge, MA}: \bibinfo{publisher}{MIT Press}.

\end{thebibliography}

\end{document}